\documentclass[useAMS,usenatbib,useAMS]{mnras}

\usepackage{newtxtext,newtxmath}

\usepackage[T1]{fontenc}

\DeclareRobustCommand{\VAN}[3]{#2}
\let\VANthebibliography\thebibliography
\def\thebibliography{\DeclareRobustCommand{\VAN}[3]{##3}\VANthebibliography}

\usepackage{graphicx}	
\usepackage{amsmath}	
\usepackage{rotating}
\usepackage{times}
\usepackage{longtable}

\title[MW halo inhabitants]{The accretion history of the Milky Way. I. How it shapes globular clusters and dwarf galaxies}

\author[Francois Hammer et al.]{
Francois Hammer$^{1}$\thanks{E-mail:francois.hammer@obspm.fr},
 Hefan Li$^{2}$$^{,}$$^{3}$,
 Gary A. Mamon$^{4}$,
 Marcel S. Pawlowski$^{5}$,
 Piercarlo Bonifacio$^{1}$,
 Yongjun Jiao$^{1}$, \newauthor
 Haifeng Wang$^{1}$$^{,}$$^{6}$,
 Jianling Wang$^{3}$,
 Yanbin Yang$^{1}$
\\
$^{1}$GEPI, Observatoire de Paris, Paris Sciences et Lettres, CNRS, Place Jules Janssen 92195, Meudon, France.\\
$^{2}$School of Physical Sciences, University of Chinese Academy of Sciences, Beijing 100049, P. R. China\\
$^{3}$CAS Key Laboratory of Optical Astronomy, National Astronomical Observatories, Beijing 100101, China\\
$^{4}$Institut d'Astrophysique de Paris (UMR7095: CNRS \& Sorbonne Universit\'e), 98 bis Bd Arago, 75014, Paris, France\\
$^{5}$Leibniz-Institut fuer Astrophysik Potsdam (AIP), An der Sternwarte 16, D-14482 Potsdam Germany\\
$^{6}$CREF, Centro Ricerche Enrico Fermi, Via Panisperna 89A, I-00184, Roma, Italy
}

\date{Accepted 2022 December 12. Received 2022 November 30; in original form 2022 September 06}

\pubyear{2022}

\begin{document}
\label{firstpage}
\pagerange{\pageref{firstpage}--\pageref{lastpage}}
\maketitle

\begin{abstract}
Halo inhabitants are individual stars, stellar streams, star and globular
clusters, and dwarf galaxies. Here we compare the two last categories that
include objects of similar stellar mass, which are often studied as
self-dynamical equilibrium systems. We discover that the half-light radius of
globular clusters depends on their orbital pericenter and total energy, 
  and that Milky Way (MW) tides may explain the observed correlation. We also
  suggest that the accretion epoch of stellar systems in the MW halo
  can be calibrated by the total orbital energy, and that such a relation is
  due to both the mass growth of the MW and dynamical friction
  affecting mostly satellites with numerous orbits. This calibration starts
from the bulge, to Kraken, Gaia Sausage Enceladus, Sagittarius stellar
systems, and finally to the new coming dwarfs, either or not linked to
  the vast-polar structure. The most eccentric globular clusters and dwarfs
  have their half-light radius scaling as the inverse of their binding
  energy, and this over more than two decades. This means that earlier
  arriving satellites are smaller due to the tidal effects of the MW. Therefore, most halo inhabitants appear to have their structural parameters shaped by MW tides and also by ram-pressure for the most recent arrivals, the dwarf galaxies. The correlations found in this study can be used as tools to further investigate the origin of globular clusters and dwarfs, as well as the assembly history of our Galaxy. 
\end{abstract}




\section{Introduction}

The stellar system content of the Milky Way ({\rm MW}) halo has been used to determine its merger history, the understanding of which has been considerably improved with the orbital constraints derived from the all-sky Gaia EDR3 observations. Since the dynamical probes of the MW gas disk are limited by 20-25 {\rm kpc} in radial distance, the halo inhabitants are also used as dynamical probes to constraint the mass of our Galaxy at large radii, and ultimately, its total mass. 

However, which type of orbital equilibrium with the MW potential has been reached by the halo stellar systems? Are their orbits ruled by the Jeans equation meaning that they can be used to probe the MW mass? Are they at pseudo equilibrium as, e.g., regularly affected by tidal shocks such as globular clusters (GCs) when passing near the inner bulge, or when crossing the disk \citep*{Aguilar1988,Gnedin1999}? Alternatively, could some of them be near their first pericentric passage having recently lost their gas and being mostly out of equilibrium \citep{Yang2014}? Comparative studies of their orbits could bring some answers to these questions.

Thanks to the Gaia astrometric mission, considerable efforts have been made to reveal the history of the ancient merger events having affected the Milky Way \citep{Kruijssen2019, Kruijssen2020,Massari2019,Malhan2022}. This has generated a large variety of events often nicknamed on the basis of various mythologies. Perhaps, the most achieved effort to date has been made by \citet{Malhan2022} who used specific search tools that compares angular momenta, total energy of 170 globular clusters, 41 stellar streams and 46 dwarf galaxies, providing some homogeneity in characterizing previous events. Apart {\rm Sagittarius} (Sgr) and Willman I, their classification of ancient mergers does not include dwarf galaxies, which are likely newcomers due to their considerably higher energy and angular momentum \citep{Hammer2021}.

Section~\ref{sec:data} presents the data from Gaia EDR3  and indicates how
they have been analyzed to produce orbital quantities, and it is accompanied
by Appendix~\ref{sec:tables} containing the result for
GCs. Section~\ref{sec:corr} reveals the strong correlation between intrinsic
($r_{\rm half}$) and orbital ($R_{\rm peri}$) parameters, and provides an
interpretation of its origin. Section~\ref{sec:E-time} suggests that the
epoch of infall of stellar systems in the MW halo can be approximated by the
total energy, which is discussed and explained in Section~\ref{subsec:time}
and implies that dwarfs are newcomers. 
In Section~\ref{sec:discussion} we also show that stellar systems (GCs and dwarfs) with large eccentricity are actually affected by MW tides, and then how dwarfs have different nature depending on whether they belong to the Vast Polar structure or not.

\section{Data and derived orbital parameters}
\label{sec:data}
The goal of this paper is to investigate whether the intrinsic properties of MW halo inhabitants are consistent with self-equilibrium or determined by their orbital properties. Despite the importance of previous theoretical investigations, few has been done on the observational side, and this paper aims at filling this gap. 

\begin{figure}
\center
\includegraphics[scale=0.24]{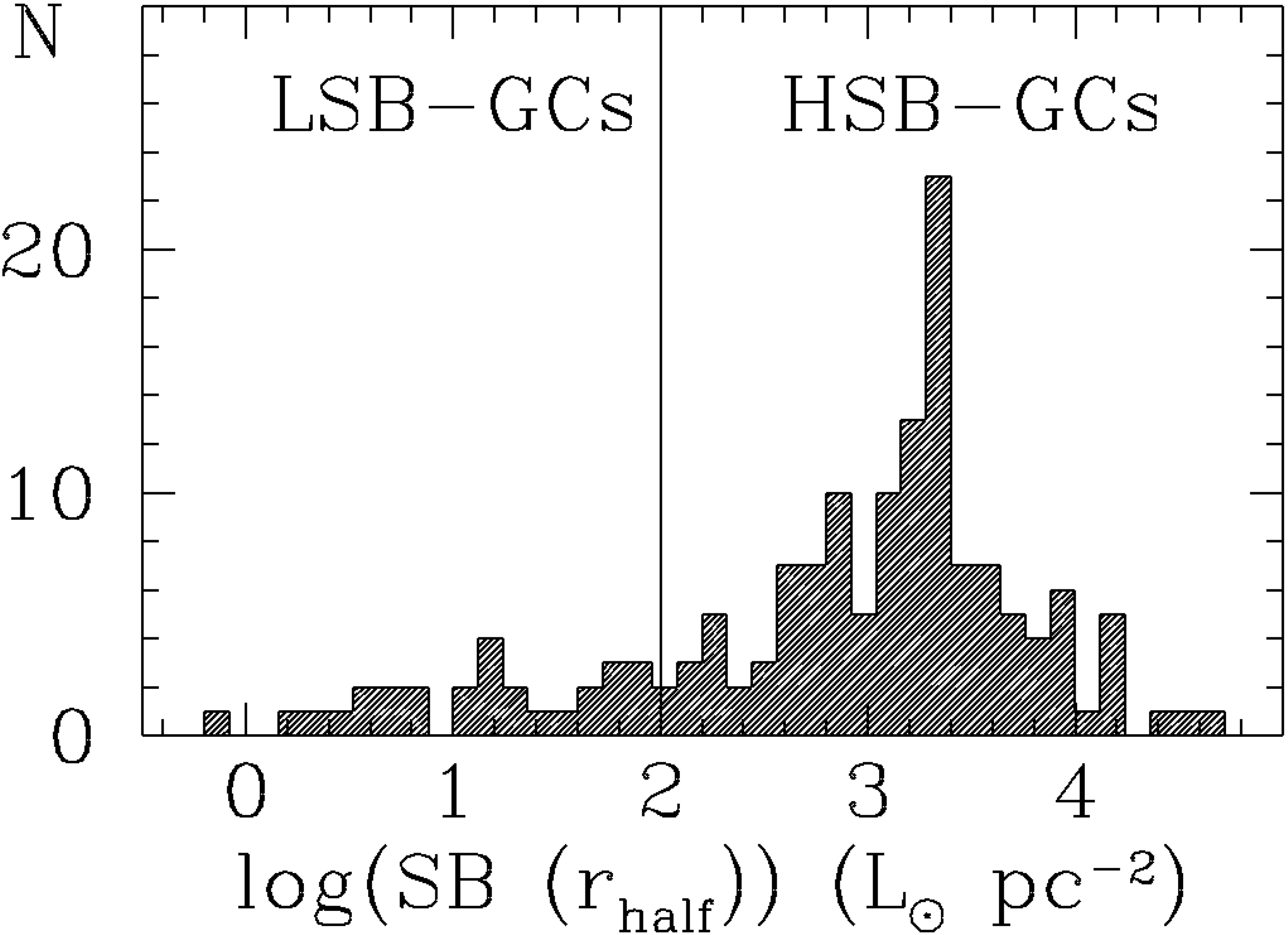}
\caption{Distribution of globular cluster surface brightness in logarithmic
  scale. Surface-brightness in the $V$ band have been averaged within the
  half-light radius. The 127 {\rm  HSB-GCs} (log(SB/$\rm L_{\odot}pc^{-2}$)
  $>2$) distribute almost along a gaussian with a peak at log({\rm SB/$\rm L_{\odot}pc^{-2}$}) = 3.3. The 29 {\rm  LSB-GCs} include the {\rm  GCs} represent the low surface brightness tail of the GC distribution, and they are on average 315 times fainter than {\rm  HSB-GCs}.}
\label{fig:histo}
\end{figure}

Here we consider all stellar systems as a single class aiming not to distinguish between globular clusters and dwarfs. Our motivation is two-fold: first, these systems have a similar range in stellar mass, and second, as pointed out by \citet{Marchi-Lasch2019}, there are similarities between structural properties of ultra faint dwarfs and low surface brightness GCs. This leads us to distinguish three different populations, the high-surface brightness ({\rm  HSB-GCs}, with log(SB/$L_{\odot} {\rm pc^{-2}}$) $>$2), the low-surface brightness (LSB-GCs, with log(SB/$L_{\odot} {\rm pc^{-2}}$) $<$2) globular clusters (see Figure~\ref{fig:histo}), and the dwarfs for which the surface brightness (SB) is similar or smaller than that of {\rm  LSB-GCs}. The classification by surface brightness is possibly more restrictive than that of \citet{Marchi-Lasch2019} in selecting {\rm  LSB-GCs}. It is also much more accurate, because it comes from the work of \citet{Baumgardt2020} who significantly improve the determination of absolute luminosity, providing corrections that can be as high as two magnitudes in $V$-band. To do so, they counted individual stars using HST photometry, selecting them from their proper motion, color-magnitude diagram and radial velocity.

In the following, we have used the data\footnote{Globular cluster data has been compiled by Holger Baumgardt, Antonio Sollima, Michael Hilker, Andrea Bellini \& Eugene Vasiliev (see https://people.smp.uq.edu.au/HolgerBaumgardt/globular/).} for 156 {\rm  GCs} from  \citet{Baumgardt2017,Baumgardt2018,Baumgardt2020,Baumgardt2021,Sollima2017}, and their proper motions from Gaia EDR3 \citep{Vasiliev2021}. Tables~\ref{tab:sph} and \ref{tab:dyn} in Appendix~\ref{sec:tables} display the resulting orbital parameters (velocities and orbital radii) and their error bars for a MW model following \citet[see also more descriptions in \citealt{Jiao2021}]{Eilers2019}. For dwarfs, we are using the data from Gaia EDR3 \citep{Li2021} using the same MW model, and same prescriptions for deriving parameters. The \citet{Eilers2019} model of the MW rotation curve includes a bulge, a thick and a thin disk following \citet{Pouliasis2017}. The halo is represented by a NFW model \citep*{Navarro1996} that requires a cut-off radius fixed at $R_{\rm vir}$= 189 {\rm kpc} to avoid infinite mass; this reduces the effect of the mass exterior to a given radius in the potential calculation by +4$\pi$G$\rho_{\rm 0}$ $R_{\rm s}^{2}$/(1+$R_{\rm vir}/R_{\rm s}$) after combining Eqs. 2.28 and 2.64 of \citet{Binney2008}, where $\rho_{\rm 0}$= 0.0106 ${\rm M_{\odot} pc^{-3}}$ is the central density of the halo and $R_{\rm s}$=14.8 {\rm kpc} is the halo scale radius. 

\section{Is there a correlation between intrinsic and orbital parameters?}
\label{sec:corr}
\subsection{Correlation between half-light radius and pericenter}

\begin{figure}
\center
\includegraphics[scale=0.28]{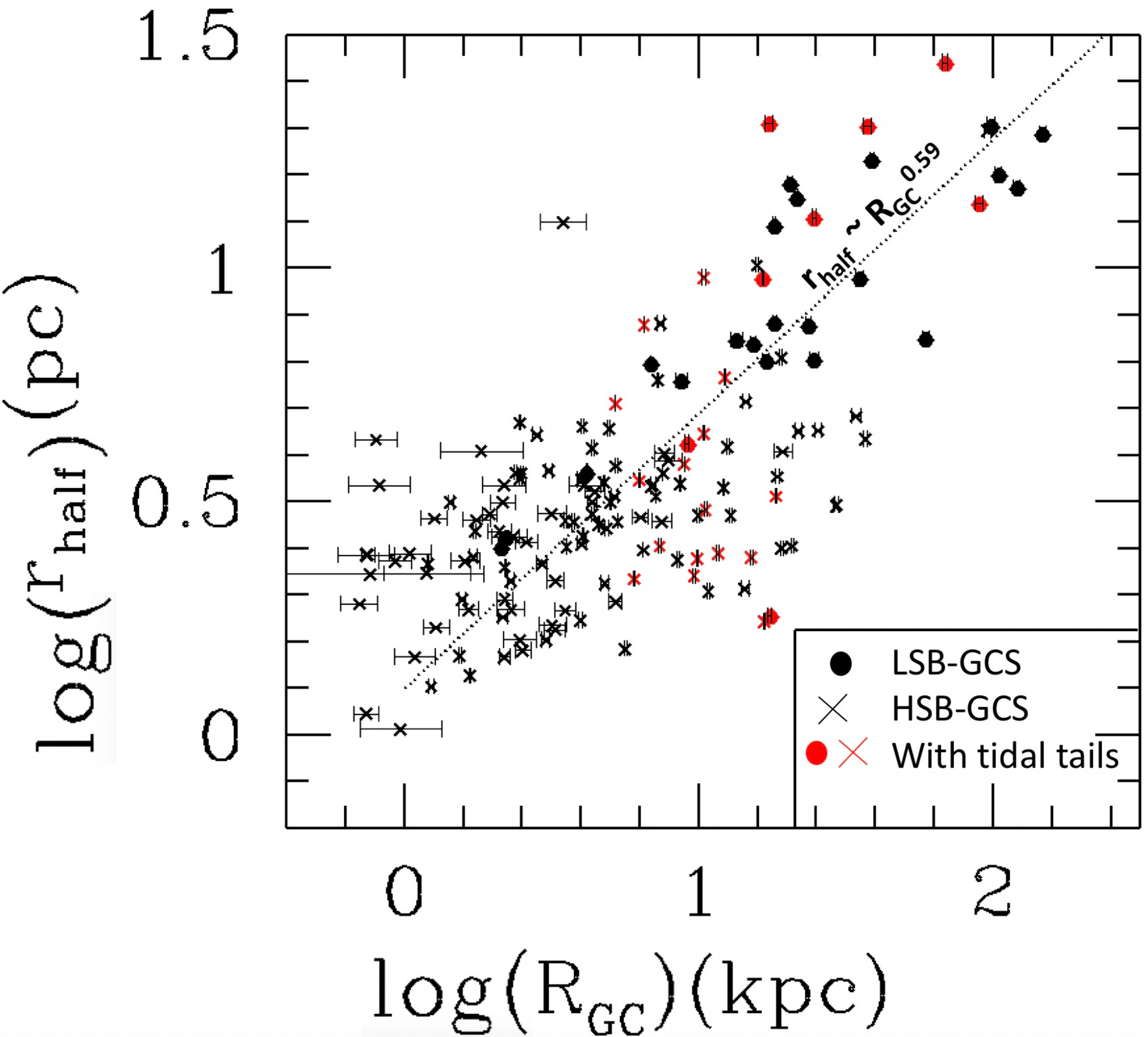}
\caption{Half-light radius versus their galactocentric distances ($R_{\rm
    GC}$) for {\rm  HSB-GCs} (\emph{crosses}) and {\rm  LSB-GCs}
  (\emph{filled circles}). \emph{Red symbols} identify {\rm  GCs} with tidal tails, which has been based on the compilation made by \citealt*{Zhang2022}.} 
\label{fig:rh_RGC}
\end{figure}

Pioneering studies of \citet{vandenBergh1994,vandenBergh2011,vandenBergh2012} revealed the existence of a correlation between GC half-light radii and their galactocentric distances, $R_{\rm GC}$. Figure~\ref{fig:rh_RGC} shows that it is still present despite significant progresses made in estimating GC distances, as well as their photometric parameters \citep{Baumgardt2017,Baumgardt2018,Baumgardt2020,Baumgardt2021,Sollima2017}. For the 156 {\rm  GCs} listed in Table~\ref{tab:sph}, it leads to a correlation\footnote{Along the manuscript we have used a Spearman's rank correlation $\rho$ that does not assume any shape for the relationship between variables; the significance and associated probability of $\rho$ have been tested using  $t$= $\rho$ $\sqrt{(n-2)/(1-\rho^2)}$, which is distributed approximately as Student's $t$ distribution with $n-$2 degrees of freedom under the null hypothesis.} with $\rho$= 0.63, and the only difference with  \citet{vandenBergh2011} is that the slope of the relation is slightly less steep (0.59 instead of 0.66).

Such a correlation prompted \citet{vandenBergh2011} to suggest that core-collapsed clusters have the tendency to be near the Galactic center. This is not unexpected from the theoretical side, since the inner {\rm  GCs} suffer stronger tidal forces from the MW. Within such a context, the most energetic GC stars may leave the system after gaining energy from tidal shocks, then lowering the system binding energy, finally leading to an adiabatic contraction when leaving the pericenter. 
 
\begin{figure}
\center
\includegraphics[scale=0.5]{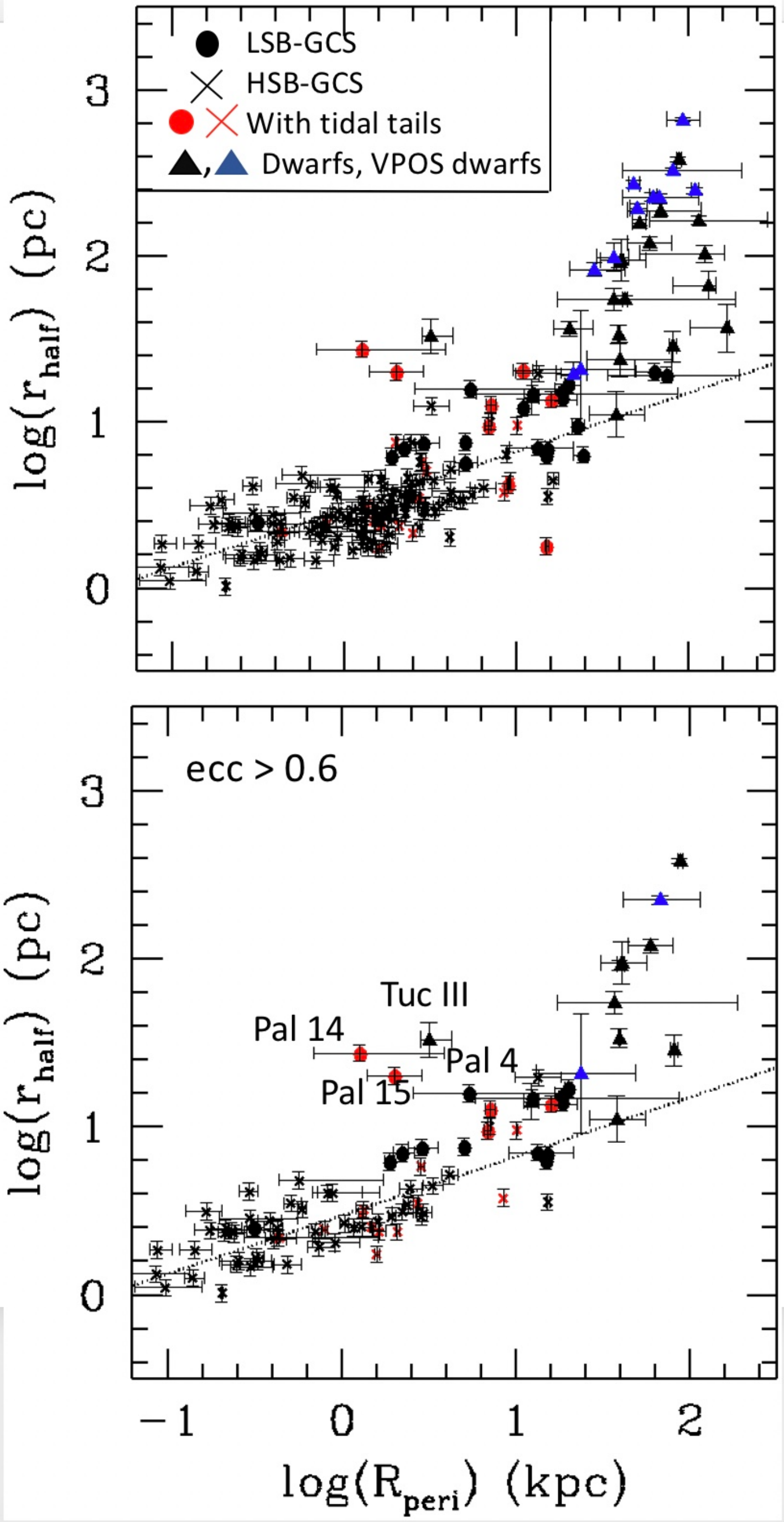}
\caption{Half-light radius versus their orbital pericenter ($R_{\rm peri}$)
  for {\rm  HSB-GCs} (crosses), {\rm  LSB-GCs} (\emph{filled circles}), and
  dwarf galaxies (\emph{triangles}). The \emph{dotted line} (slope = 0.33)
  indicates the correlation for the 156 GCs, which are shown in the \emph{top
    panel} together with 30 dwarfs, while in the \emph{bottom panel} only the 77 {\rm  GCs} and 12 dwarfs with orbital eccentricity larger than 0.6. Few stellar systems very offset from the correlation and suspected to be strongly tidally disrupted are labelled.} 
\label{fig:rh_rp}
\end{figure}

 Since the half-light radius is a fundamental structural parameter, we investigate its dependency onto a pure orbital parameter specifically the pericenter, $R_{\rm peri}$. This has been firstly investigated on the basis of Gaia DR2 data by \citet{deBoer2019}, but the error bars on pericenters were too large for providing robust conclusions. Using Gaia EDR3 data, we find a stronger correlation when replacing the galactocentric distance by the pericenter. Figure~\ref{fig:rh_rp} reveals that the half-light radius of 156 {\rm  GCs} correlates well with their orbital pericenters, with  $r_{\rm half}$ following  $R_{\rm peri}^{1/3}$ with $\rho$= 0.70, $t$=12, with an extremely low probability that it occurs by chance, $P$= 1.3 $\times$ $10^{-23}$.

Few preliminary remarks can be made about the correlation shown in Figure~\ref{fig:rh_rp}, which are particularly relevant when considering objects with different eccentricity, which we define as:
\begin{equation}
  {\rm ecc}=\frac{R_{\rm apo}-R_{\rm peri}}{R_{\rm apo}+R_{\rm peri}} \ .
  \label{ecc}
  \end{equation}

\begin{itemize}
\item The correlation becomes even more significant ($\rho$= 0.79 and $P$= 9 $\times$ $10^{-18}$) for half of the sample, i.e. the 77 {\rm  GCs} with high eccentricity (ecc$>$ 0.6, see bottom panel of Figure~\ref{fig:rh_rp});
\item The above is true even after accounting for the few {\rm  GCs} that are fully off from the correlation, especially Pal 14, Pal 15 and Pal 4, and which are all with very eccentric orbits, while two of them possess tidal tails (see red dots);  
\item  Dwarf galaxies from  \citet{Li2021} show a much steeper slope but with a less tight correlation ($\rho$= 0.55, $P$= 2 $\times$ $10^{-4}$ for 37 dwarfs).
\end{itemize}

The significant correlation shown in Figure~\ref{fig:rh_rp} suggests that MW tides could be at work, and that they are even more efficient for {\rm  GCs} with eccentric orbits. The strongest tidal shocks on {\rm  GCs} are generated as the cluster crosses the galactic disk or passes through pericenter on a highly eccentric orbit \citep{Binney2008}. However, one may wonder whether the observed correlation could be a hard-to-disentangle combination of star formation processes and subsequent dynamical evolution in the hierarchical build-up of the Milky Way halo. Former studies have indicated how the MW may have filtered the different populations of stellar systems to establish the present-day GC population \citep{Fall1977,Fall1985,Gnedin1997}. In particular, \citet[see their Figure 4]{Fall1985} predicted that the combination of tides and thermal instabilities may lead to a decreasing density with the galactocentric distance, which seems quite consistent with the ($r_{\rm half}$, $R_{\rm GC}$) correlation shown in Figure~\ref{fig:rh_RGC}, and then with the ($r_{\rm half}$, $R_{\rm peri}$) correlation shown in Figure~\ref{fig:rh_rp}.

 A major drawback with this possibility is that the \citet{Fall1985} scenario assumed a monolithic collapse for the initial formation of the MW. In the recent literature the favored formation scenario
for spiral galaxies is through the hierarchical scenario, i.e., mostly through gas-rich major mergers \citep{Hammer2005,Yang2008}, which are essential in providing enough angular momentum into galactic disks \citep{Hammer2009,Hopkins2009,Hopkins2010,Stewart2009}. This also applies to the MW, as evidenced by the finding of GSE imprints in the Energy-Angular momentum diagram \citep[see \citealt{Malhan2022} for imprints related to GCs]{Haywood2018,Belokurov2018}. If GCs were formed initially as proposed by \citet{Fall1977} and \citet{Fall1985}, their density-galactocentric distance relation would be rapidly diluted  after one or more major merger. 

 \subsection{Is the ($r_{\rm half}$, $R_{\rm peri}$) correlation due to MW
   tides?}
 Because $r_{\rm half}$ is much more correlated to $R_{\rm peri}$ than to $R_{\rm GC}$, one may conclude that (1) the correlation shown in Figure~\ref{fig:rh_RGC} is generated by that in Figure~\ref{fig:rh_rp}, and (2) it is a strong indication that GC structural parameters are impacted by pericenter passages. Simulations are necessary to go beyond that understanding, and we have used those made by \citet{Martinez-Medina2022}, who tested how the GC half-light radii with different concentration parameter values can be affected by pericenter passages. Interestingly, they found that half-light radii of concentrated GCs have shrunk by $\sim$ 3\% after one pericenter passage, while conversely, the least concentrated GCs show an expansion of their half-light radii by $\sim$ 7\%.  In the following, we investigate a very simple model based on the \citet{Martinez-Medina2022}'s simulations, and accordingly to the latter, we have considered equal the impact of tidal shocks due to a point mass and to disk passages.
 
\begin{figure}
\center
\includegraphics[scale=0.25]{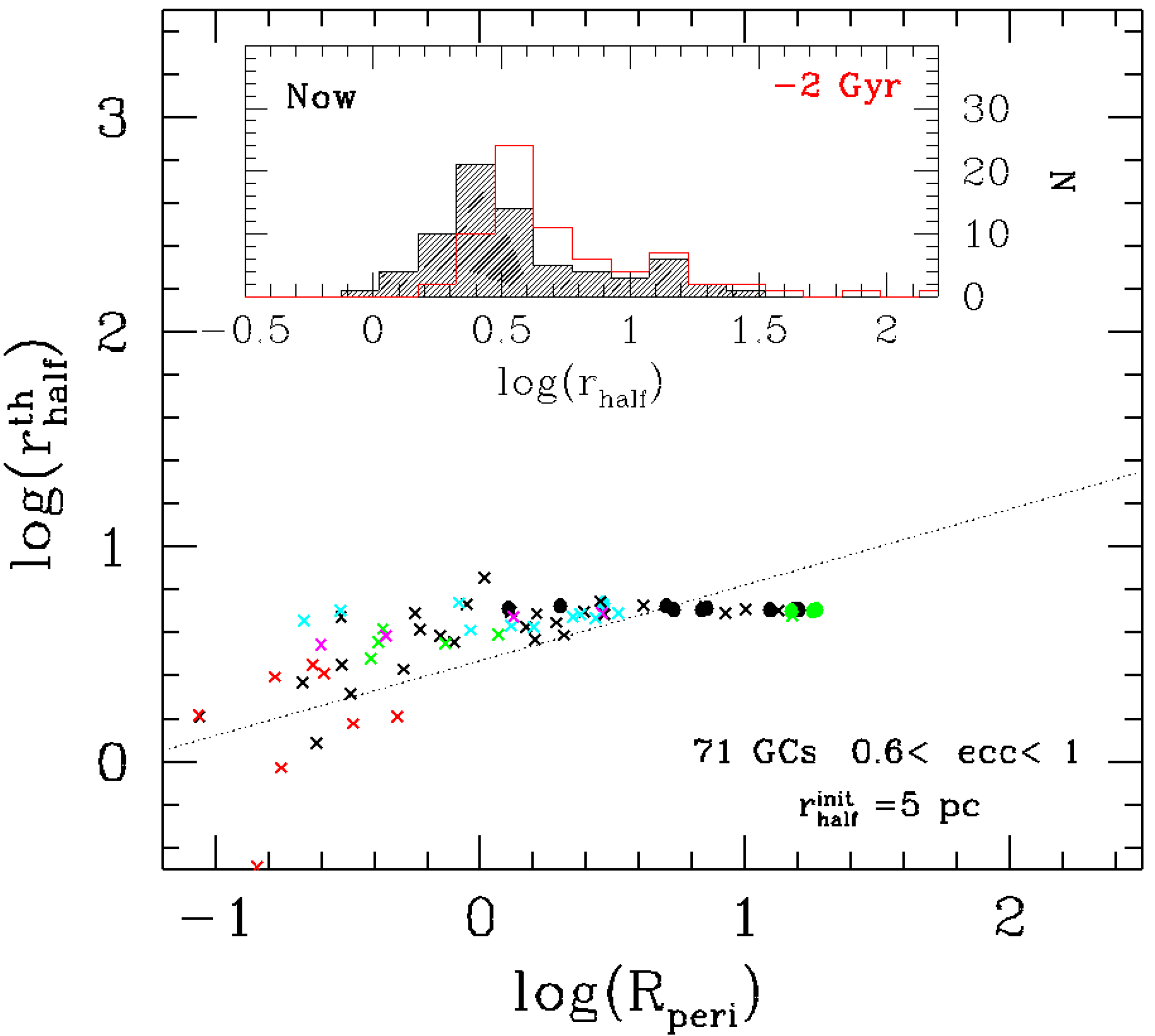}
\caption{Predicted half-light radius (in parsec) versus orbital pericenter
  ($R_{\rm peri}$, in kpc) for {\rm  HSB-GCs} (\emph{crosses}) and {\rm
    LSB-GCs} (\emph{filled circles}) after 2 Gyr evolution of GCs that were assumed having an initial half-light radius of 5 pc. The dotted line (slope= 0.33) indicates the correlation for the 156 GCs defined in Figure~\ref{fig:rh_rp}.  The \emph{inset} shows the histograms of present-day GC $r_{\rm half}$ at present (\emph{black}) and 2 Gyr ago (\emph{red}).} 
\label{fig:rh_evolution}
\end{figure}

We have verified from the \citet{Harris2010} compilation of 144 GCs that HSB-GCs have an average concentration parameter of $W_{\rm 0}$ =7.2, which is quite similar to that of compact clusters chosen by  \citet[$W_{\rm 0}$ =8]{Martinez-Medina2022}. We have further assumed a linear interpolation of the half-light radius evolution with pericentric passage (see details in Appendix~\ref{sec:cSB}), from $W_{\rm 0}$ =8 GCs ($\Delta$$r_{\rm half}$/$r_{\rm half}$= -3\%) to $W_{\rm 0}$ =2 GCs ($\Delta$$r_{\rm half}$/$r_{\rm half}$= +7\%).  To test how tidal shocks may explain the ($r_{\rm half}$, $R_{\rm peri}$) correlation, we have adopted a single initial half-light radius for all GCs to verify how it evolves in Figure~\ref{fig:rh_evolution}.  The latter shows that the correlation is mostly retrieved including its significance ($\rho$= 0.76) and its slope (0.39).  It indicates that the modest loss of half-light radius at each pericenter is compensated by the numerous pericenter passages (up to 100 for a 2 Gyr duration) experienced by the closest GCs to the center.

The evolution of GC half-light radii is also illustrated in the inset of
Figure~\ref{fig:rh_evolution}. Assuming the present-day GC $r_{\rm half}$ and
an evolution driven by tidal shocks, one finds that the peak of the
distribution evolves to $r_{\rm half}$= 2.5 pc nowadays from 3.5 pc 2 Gyr
ago. One may also notice that the correlation is better retrieved after
considering eccentric GCs. When including the less eccentric GCs (ecc $< $ 0.6), the correlation significance and the slope drop to 0.54 and 0.22, respectively. We have also tried larger look back times, such as e.g., 6 instead of 2 Gyr ago. It has led to larger slopes, which are due to extremely small theoretical GCs due to their extremely large number of pericenter passages. The above results show the limits of our exercise: the most eccentric GCs are expected to be the most affected in the last 2 Gyr, while extending our test to much larger elapsed periods leads to unrealistic values for $r_{\rm half}$ since the pericenter is likely evolving with time.

We also notice that the points with $\log(R_{\rm peri}) > 0.5$ in
Figure~\ref{fig:rh_evolution} do not follow the correlation, and that the
corresponding GCs have essentially kept their initial $r_{\rm half}$
value. They are mostly LSB-GCs, and we find that they have experienced only a
single pericenter passage during the 2 Gyr elapsed time, i.e., they have not
yet been reshaped by tides. Given that actual LSB-GC (as well as dwarf)
half-light radii are mostly above the $r_{\rm half}$ $\sim$ $R_{\rm
  peri}^{1/3}$ line in Figure~\ref{fig:rh_rp}, this indicates that the their
structural parameters are mostly fixed by their star formation processes, or
alternatively by another mechanism (see section~\ref{sec:discussion}). Our
model is surely over-simplified, and  a more sophisticated model is
necessary, though beyond the scope of the present paper. Such a more detailed
study should account for the full evolution of GCs, including their orbital
history, as well as for the MW mass evolution, and this would very useful for
disentangling the different effects on half-mass radii from tides and from star formation. On the other hand, it is unlikely that the present-day distribution of GC half-light radii can be attributed to their initial distribution function, while accounting for existing simulations \citep{Martinez-Medina2022} provides sufficient evidence that GCs are shaped by MW tides.

Figures~\ref{fig:rh_rp} and \ref{fig:rh_evolution} show that at present-day,
eccentric GCs with small pericenters are the most impacted by tides. Appendix~\ref{sec:cSB} details how the correlation between GC surface brightness and
concentration parameter has been established. We have also determined that
the few {\rm  GCs} (especially Pal 14, Pal 15 and Pal 4) that escape the
correlation shown in Figure~\ref{fig:rh_rp}  are heavily tidally stripped or shocked at present-time (see Appendix~\ref{sec:tides}).


\section{Calibrating the epoch of GC emergences in the MW halo}
\label{sec:E-time}

The correlation between an intrinsic quantity,  the half-light radius, with an orbital parameter, the pericenter, is suggestive of the important role played by the MW potential in shaping GCs. However, to compare its impact on the various inhabitants of the halo, we need to calibrate their relative epochs of infall, or alternatively, the time when they have emerged into the MW halo. This can be approximated from numerical simulations \citep{Boylan-Kolchin2013}, which show that the total energy is decreasing with infall look-back time, similarly to the onion skin model initiated by \citet{Gott1975} for explaining the outer density profile of elliptical galaxies. This does not come at a surprise, since the more orbits made by a stellar system in the halo, the more dissipated its energy would be\footnote{This also applies to angular momentum or to pericenter, both quantities that should decreases with time; however, small pericenter or angular momentum values, can be also associated to radial orbits, and hence these two quantities cannot be taken as time proxy.}, e.g., by encountering other sub-systems including giant molecular clouds, or at passages through the MW disk, or near the bulge. 

We estimate the epochs of infall with the help of the identifications of substructures in the Galactic halo made by \citet{Malhan2022}, which have been assumed to be linked to past merger events. By studying the associations of these events with GCs, \citet{Malhan2022} have been able to identify substructures associated to the bulge, to the disk, to a novel substructure called {\rm Pontus}, and also to {\rm Gaia Sausage Enceladus} ({\rm GSE}), LMS1-Wukong, and  Sgr. To this we have added the {\rm Kraken} event identified by \citet{Kruijssen2020} because it could be the earliest merger that can be identified in the MW halo. We also notice that \citet{Kruijssen2020} associated to {\rm GSE} some of the {\rm  GCs} associated to {\rm Pontus} by \citet{Malhan2022}.  

\begin{figure*}
\includegraphics[width=6.4in]{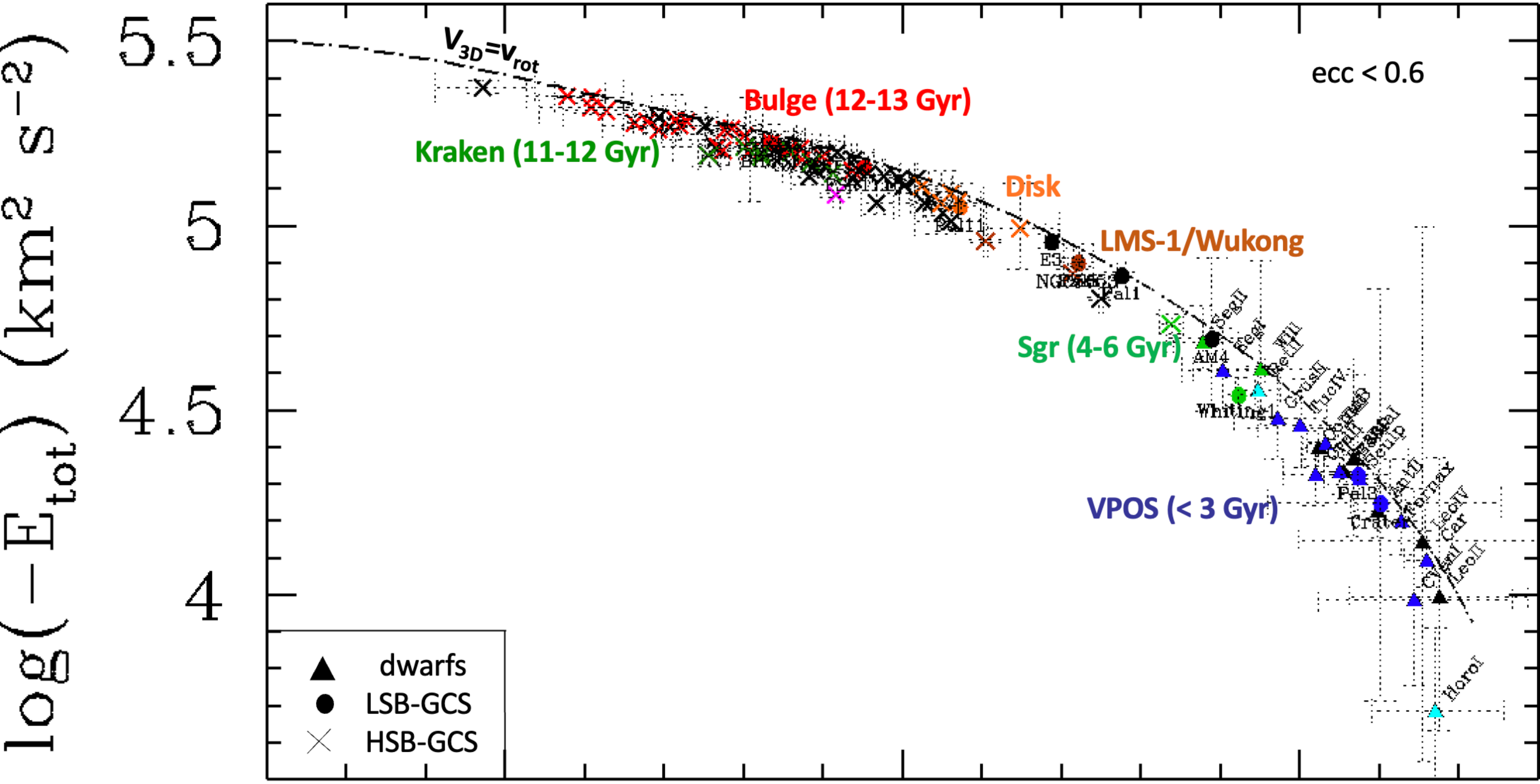}
\includegraphics[width=6.4in]{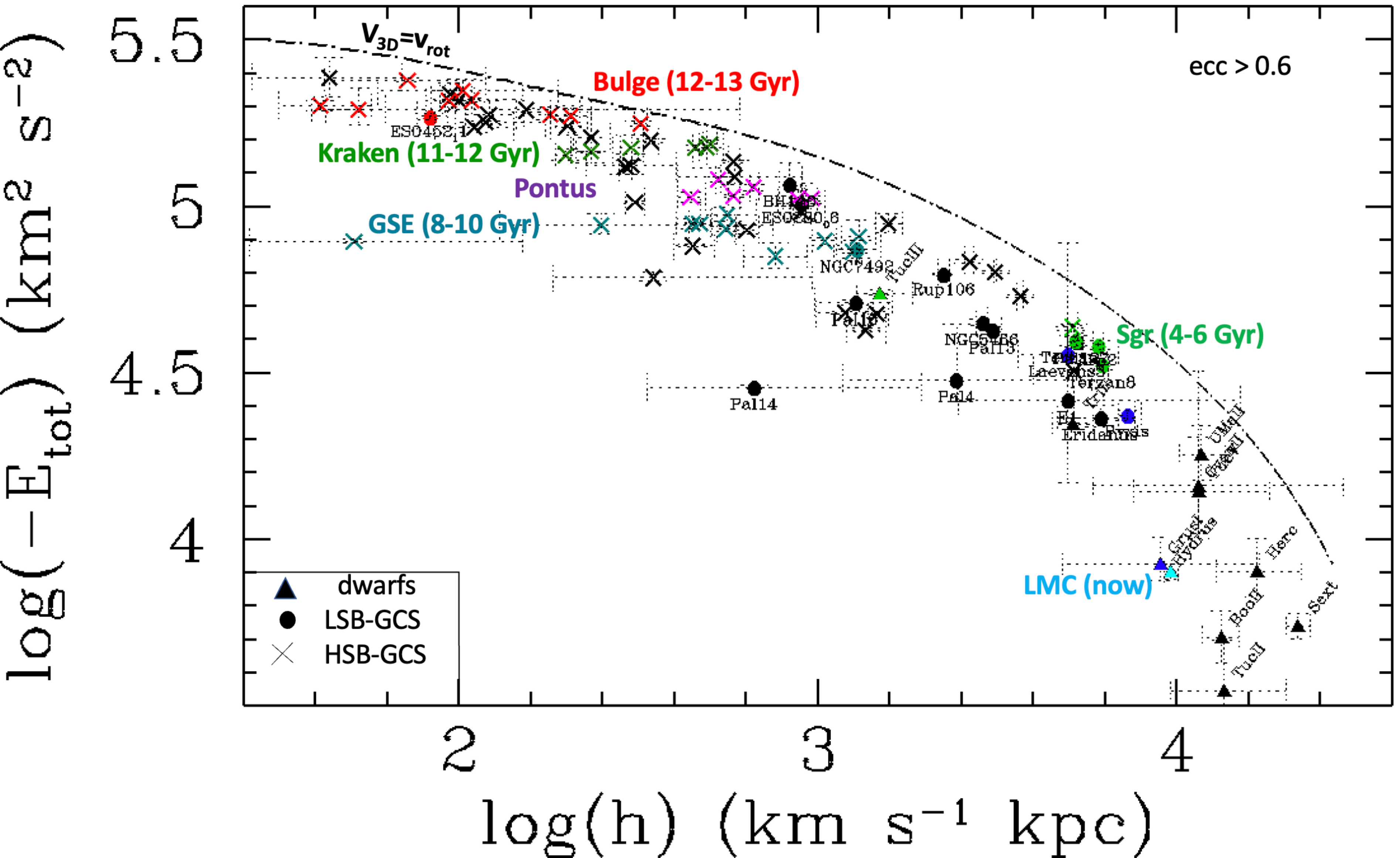}
\caption{Total energy versus the angular momentum, $h$=$R_{\rm GC}$ $\times$
  $V_{\rm tan}$ in logarithmic scale, for {\rm  HSB-GCs} (crosses), {\rm
    LSB-GCs} (\emph{filled circles}), and dwarfs (\emph{triangles}). Identified structures by
  \citealt{Malhan2022} and \citealt{Kruijssen2020} are added by different
  colors indicated in the Figure. {\rm  VPOS}  dwarfs and {\rm  GCs} are also
  shown in \emph{blue}, while the few dwarfs associated to the LMC are shown
  with \emph{cyan}. The \emph{top panel} shows 79 {\rm  GCs} and 20 dwarfs with
  eccentricities smaller than 0.6, while the \emph{bottom panel} shows the 77 {\rm
    GCs} and 15 dwarfs with eccentricities larger than 0.6. The \emph{dotted line}
  shows the limit that cannot be passed by any orbits, as it is fixed for
  $V_{\rm rad}$=0 and a circular orbit (lowest possible energy for a given
  angular momentum). Note that few dwarfs having a positive energy are
    not represented, including Leo I in the {\rm VPOS} , Carina II and III
    and Phoenix II, all related to the {\rm LMC}, and Hydra II and Leo V.
 }
\label{fig:E-h-ecc}
\end{figure*}

 
Figure~\ref{fig:E-h-ecc} presents the relation between the total energy and the angular momentum for eccentric (bottom) and non-eccentric (top) stellar system. First, it is very similar to that between the total energy and the pericenter, simply because the angular momentum correlates extremely well with the pericenter. Second, the total energy appears to be well associated to the time occurrence of the merger events. In most halo substructures, {\rm  GCs} show a narrow range in energy  suggesting that the total energy is a proxy of the infalling time.

If correct, it would assume a very early epoch for the  bulge {\rm  GCs} as
indicated by their ages \citep{Kruijssen2019}. This is indeed corroborated by
the fact that most bulge {\rm  GCs} have their orbits circularized, as
indicated by the comparison between the two panels of
Figure~\ref{fig:E-h-ecc} showing that most bulge {\rm  GCs} have low
eccentricity orbits. We have followed \citet{Kruijssen2020} by
  indicating {\rm Kraken} as being the most ancient detected merger in the MW
  halo, this followed by {\rm Pontus}, and that by {\rm GSE} with an
  approximated age of 8-10 Gyrs, and for which ages have been derived from
  that of their associated {\rm  GCs} \citep{Kruijssen2019}.  The methodology
  of \citet{Kruijssen2020} is to compare the observed GC age distribution
  from \citet{Kruijssen2019} to that derived from cosmological
  simulations. They associate a merger with an epoch defined as  ``as the
  moment at which SUBFIND can no longer find a bound subhalo, and the subhalo
  is therefore considered to have merged into the halo of the central
  halo''. In principle, this corresponds to the time when the merger is close
  to completion, though it appears quite uncertain since, in their subsequent work,
  \citet{Kruijssen2020} assumed instead that it corresponds to the entrance
  of the system into the halo. On the other hand, during merging encounters,
  star formation usually reaches its peak at the time of merger \citep[see their fig. 4]{Puech2012} rather than
  at first pericenter \citep{DiMatteo2008}. Such star formation at merging
   is likely
  accompanied with the formation of proto-GCs. 

However, the above uncertainty has small consequences in the case of major
merger events such as Kraken or GSE, because due to dynamical friction, the
time between halo entrance and final merger is relatively small and similar
to their quoted uncertainties, i.e., $\Delta T  \sim$ 2 Gyr. The situation
for Sgr is very different, because (1) the merger is far from being
completed, and (2), it is likely a minor merger. The Sgr merger event has
been considerably studied and modeled, since conversely to former merger
events, the Sgr core, associated GCs, and stream are still easily
recognizable\footnote{Besides M\,54, Arp 2, Terzan 7, Terzan 8, Whiting 1,
  Pal 12 that are part of the Sgr system \citep{Bellazzini2020}, we have
  quoted Segue II, Willman I, and Tucana III as potential companions of Sgr
  because their poles, energy and angular momentum are either similar, or
  with an inverted angular momentum; for example, in the current potential
  used in this paper, Segue II appear to have encountered Sgr less than one
  hundred million years ago.}.  It is likely that many Sgr GCs have taken
their birth into the Sgr surrounding material, i.e., before its entrance into
the MW halo. Because we are interested in the location of Sgr GCs into the
($E_{\rm tot}$, $h$) diagram, their total energy should be quite similar to that at the Sgr infall time, or its first entrance into the halo.

To estimate the Sgr infall time, we have used the most accurate analyses of the star formation history ({\rm SFH}) of Sgr. We further assume that before entering the MW halo, the Sgr progenitor should have been considerably gas-rich as suggested by the distribution of gas-rich dwarfs in the Local Group \citep{Grcevich2009}. One can then estimate its first infall and gas depletion times using the Sgr {\rm SFH}.  
Sgr hosts populations of different ages as seen in deep HST photometry \citep{Siegel2007} that
clearly displays several turnoffs in their color-magnitude diagrams. 
From the photometry and
detailed abundances for a sample of giant stars within
$9'$ from the center of M54, \citet{Mucciarelli2017} found
a double-peaked metallicity distribution function, with peaks
at $-1.5$ and $-0.5$. They also claimed
that the chemical evolution of Sgr implies a strong gas
loss occurring between  2.5 Gyr and 7.5 Gyr ago, which they 
presume occurring at the first passage at pericenter  of
Sgr following its infall into the Milky Way.
More recently, \citet{Alfaro-Cuello2019} using deep, albeit low resolution, 
spectroscopic data that covers only the inner region
of M\,54 ($2\farcm$) radius, to be compared to $7\farcm4$ of the tidal
radius of the cluster) claimed the detection of three different
populations  a young one (2.3 Gyr old), an old population
(12 Gyr old), and an intermediate-age population (4.3 Gyr old) that
is more widespread. The young population can coincide with the
last event of gas depletion that probably occurs in the very central
Sgr region, while the intermediate-age population is suggestive of
a longer event during which ram-pressure progressively removes the
gas from Sgr, while star formation is sustained by gas pressurization from $\sim$ 6 to 3 Gyr ago (see figure 7 of  \citealt{Alfaro-Cuello2019}).

 Using SDSS photometry and spectroscopy \citet*{deBoer2015} found that the {\rm SFH} of the Sgr Stream and found that the {\rm SFH} increases 6 Gyr ago with a peak 5 Gyr ago and then  
a rapid decline 4 Gyr ago (see their Figure 6). 
Interestingly, the most recent models of Sgr \citep{Vasiliev2021b,Wang2022a} assumed an infall from 3 to 4.7 Gyr ago. Finally an infall more recent than 6 Gyr ago is necessary to account for the age of Whiting 1 (6.5 Gyr old). Putting all the available evidence together we conclude that
the first infall of Sgr must have occurred between 4 and 6 Gyr ago.

Besides GCs, Figure~\ref{fig:E-h-ecc} also shows how other dwarfs than Sgr are distributed in the ($E_{\rm tot}$, $h$) plane (see also \citealt{Hammer2021}). Most of the dwarfs belonging to the Vast Polar Structure perpendicular to the disk ({\rm  VPOS}, see \citealt{Pawlowski2014,Pawlowski2018,Li2021}) show small eccentricities and follow the line with $V_{\rm rad}$=0 (or $V_{\rm 3D}$= $V_{\rm rot}$), i.e., a minimal energy for a given angular momentum (see blue triangles). This has been attributed by \citet{Hammer2021} to the fact that most of them are newcomers and have recently lost their gas due to ram-pressure, which affects more their radial velocities, leading to the observed anisotropic distribution of velocities \citep{Cautun2017,Li2021}. However, there are other dwarfs showing high energy and high orbital eccentricity, including dwarfs associated to the LMC on the basis of their relative Gaia motions \citep{Erkal2020,Patel2020}.  Because of that, we have considered two substructures of the {\rm  VPOS}, one still called the {\rm  VPOS}  (blue triangles), the other called LMC for associated dwarfs to this galaxy (cyan triangles)\footnote{It includes, Carina II, Carina III, and Phoenix II, all with positive energy for that MW mass model (not represented in Figure~\ref{fig:E-h-ecc}), and also Horologium I, Hydrus I, and Reticulum II.}. Dwarfs that are not in the {\rm  VPOS}  are shown in Figure~\ref{fig:E-h-ecc} as black triangles, and many of them have significant radial velocities, as indicated by their offset from the $V_{\rm 3D}$= $V_{\rm rot}$ line.


\section{Discussion}
\label{sec:discussion}

\subsection{Relating total energy to the epoch when stellar systems entered  the MW halo}
\label{subsec:time}

\begin{figure}
  \centering
\includegraphics[scale=0.28]{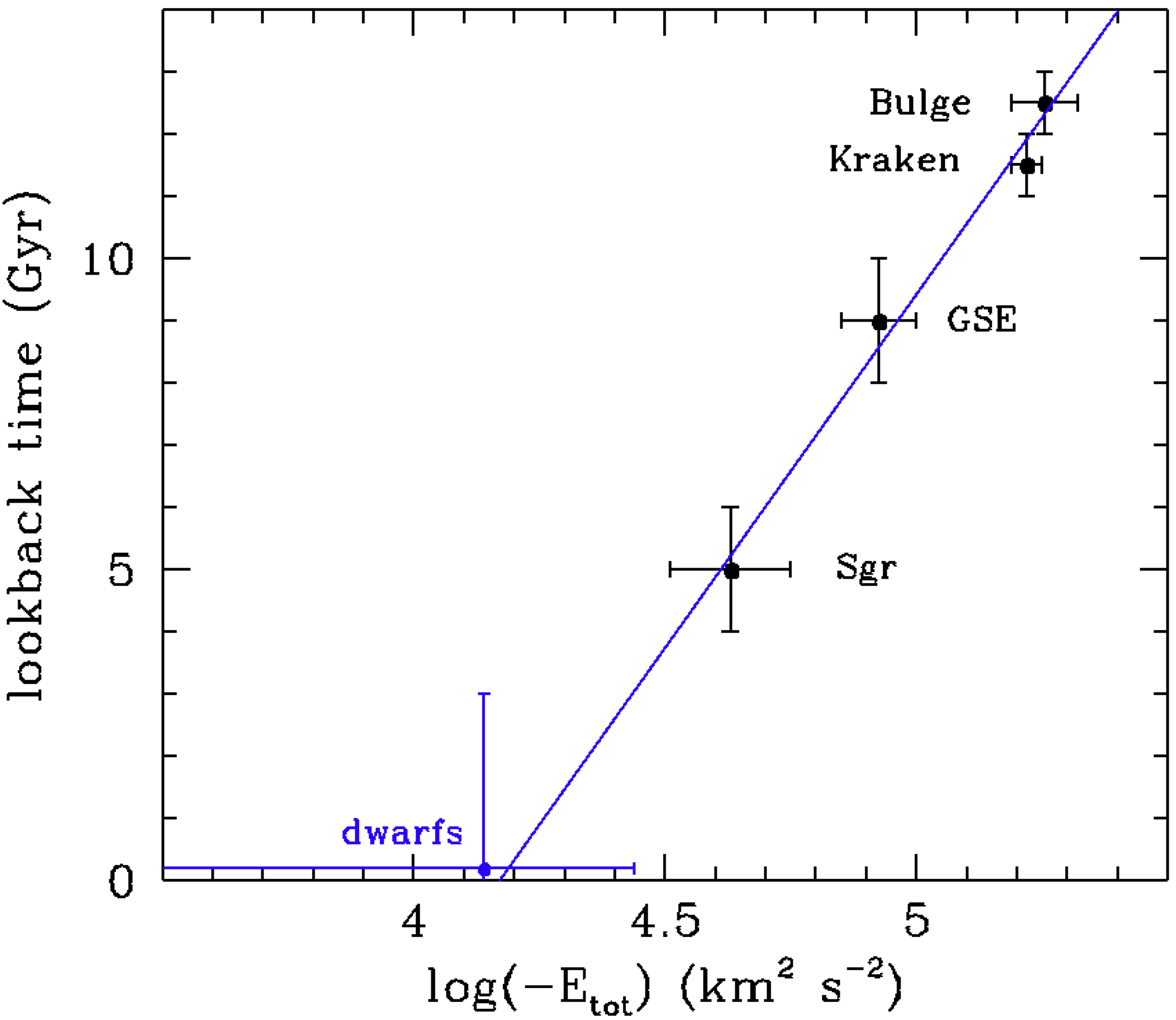}
  \caption{
Lookback time of stellar system entry in the Milky Way halo as a function of current binding
energy for different families of globular clusters, and for the 25 dwarf galaxies that do not belong to the low energetic Sgr system (excluding then Sgr, Segue II, Tucana III and Willman I) or to the high energetic LMC system (Carina II, Carina III, Horologium I, Hydrus I, Phoenix II, and Reticulum II).
The \emph{blue line} is a quick linear fit. A na\"{\i}ve interpretation is that dwarf
satellites with $\log(-E_{\rm tot}/[\rm km^2\,s^{-2}]) < 4.17$ are on their
initial approach, a value very close to the logarithm of the average energy
(4.14) of 25 dwarfs  (see text), whose scatter provides an upper limit of
$-E_{\rm tot}$ = 4.34. The latter combined with the linear fit suggests a
lookback time of halo entry smaller than 3 Gyr. 
  }
\label{fig:tvsE}
   \end{figure}

The estimated lookback times of entry into the MW halo of different
populations of GCs of \cite{Kruijssen2019} can be compared to their orbital
binding energies  (see Figure~\ref{fig:E-h-ecc} and section~\ref{sec:E-time}). This is shown in 
Figure~\ref{fig:tvsE}, which has been built on the basis of the narrow total
energy range shown by each identified structures in the MW halo.

Two simple mechanisms could explain such a relation for which stellar systems coming the earlier in the MW halo have smaller energy :
\begin{itemize}
\item The MW mass increase with time, which imposes a low energy for early-coming stellar systems;
\item Early-coming stellar systems have experienced several orbits and then have large chances to encounter other systems to lose further their energy.
\end{itemize}

Figure~\ref{fig:E-h-ecc} shows the distribution of previous mergers in the MW, illustrating also that dwarfs possess higher total energy and angular momentum than Sgr, and that of most GCs. The time elapsed between {\rm GSE} and Sgr events is $\sim$ 4 Gyr, corresponding to a difference of energy of 0.25 dex (a factor $\sim$ 1.8). Extrapolating the same energy difference towards higher energies, would lead to an energy value right at the average value for dwarfs as indicated in Figure~\ref{fig:tvsE} . Even if the total energy cannot be taken as a very precise clock, it suggests that most dwarfs are new or recent ($<$ 3 Gyr) comers into the MW halo (see also \citealt{Hammer2021}). Another clue for that is the very large scatter presented by dwarfs in the  ($E_{\rm tot}$, $h$) plane. This is because newcomers may have different origins and then possess different initial energies and angular momenta.



This contradicts several studies making the assumption that many dwarfs are
long-lived satellites of the MW, for nearly a Hubble time. Such an assumption
neglects, e.g., the impact of the MW mass evolution. For example, in
$\Lambda$CDM cosmology, the mass growth of the Milky Way since the epoch when
bulge GCs were accreted ($z=4.5$) is roughly a factor 50 (from fig.~5 of \citealt{vandenBosch02}). 
With such a small MW mass, only bulge GCs with the largest binding energies
could be bound, while it is unlikely that low-binding energy dwarfs that lie at large distances can be captured at the same time.

\begin{figure*}
\includegraphics[width=6.5in]{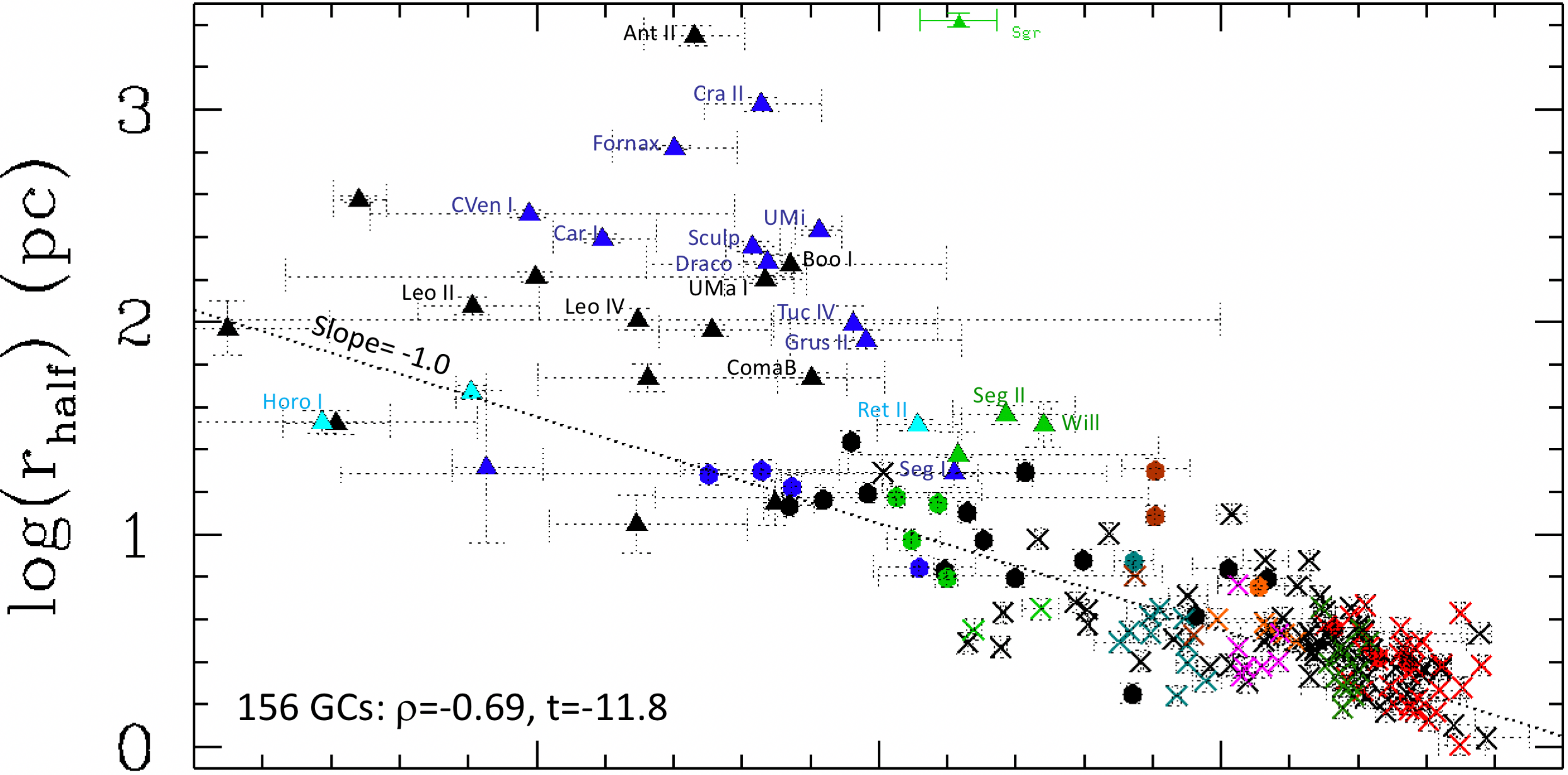}
\includegraphics[width=6.5in]{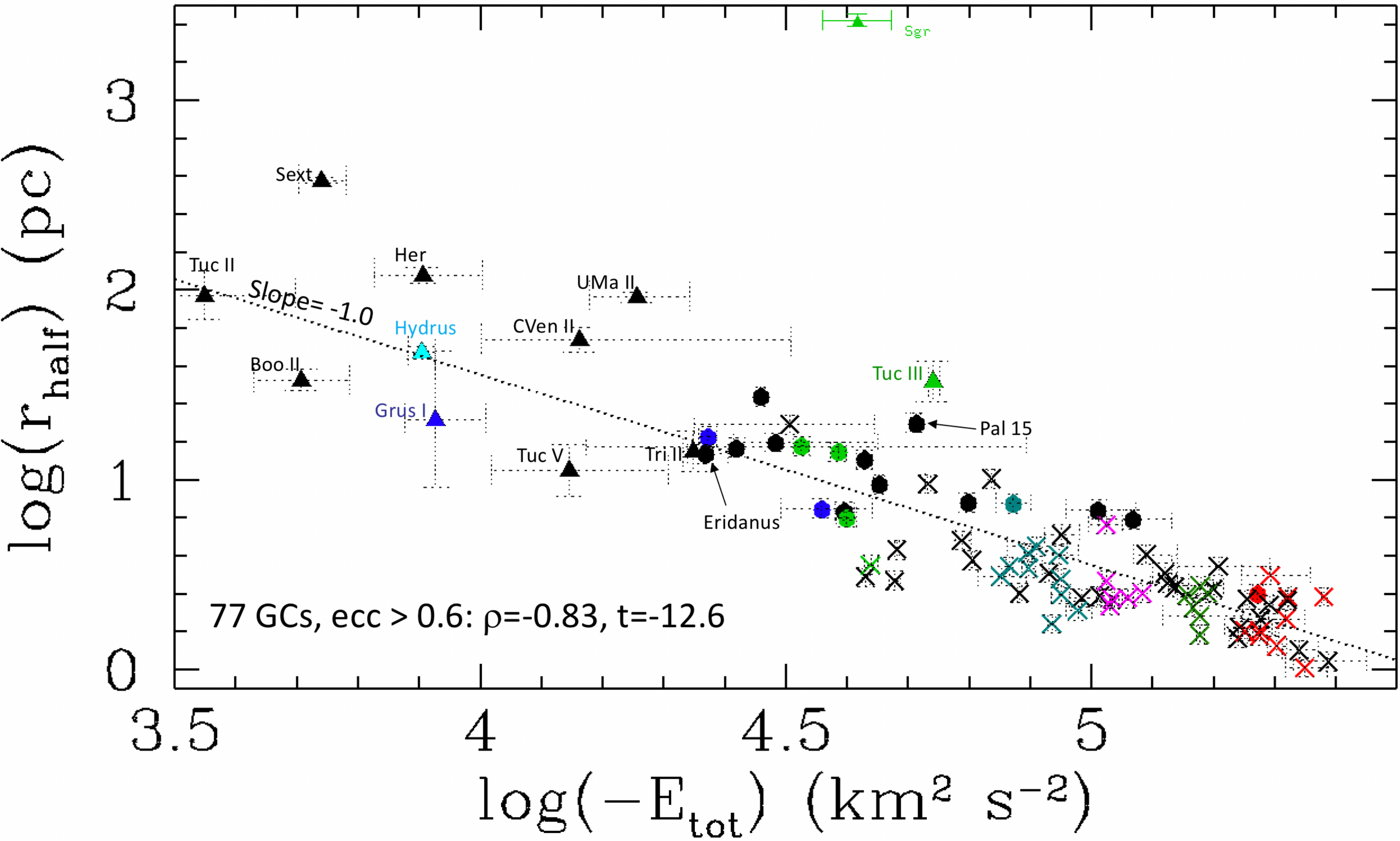}
\caption{Total energy versus half-light radius in logarithmic scale, for {\rm  HSB-GCs} (crosses),{\rm  LSB-GCs} (full dots), and dwarfs (triangles). 
 Identified structures by \citealt{Malhan2022} and \citealt{Kruijssen2020} are represented by the color codes used in Figure~\ref{fig:E-h-ecc}.
 {\rm  VPOS}  dwarfs and {\rm  GCs} are also shown in blue color, while the few dwarfs associated to the LMC are shown with cyan color. The top panel shows 156 {\rm  GCs} and 35 dwarfs (those with ecc $<$ 0.6 are labelled), while the bottom panel shows the 77 {\rm  GCs} and 15 (labelled) dwarfs with eccentricities larger than 0.6. The dotted line shows the fit of the GC points in the top panel, and of both {\rm  GCs} and dwarfs in the bottom panel, and the significance of the correlation is given on the bottom-left of each panel. } 
\label{fig:E-rh}
\end{figure*}

\subsection{The evolution of half-light radius with total energy and look-back time}
\label{sec:evolution}

The top panel of Figure~\ref{fig:E-rh} shows how the half-light radius of the 156 {\rm  GCs} correlates with the total energy, i.e., the smaller the energy, the smaller the half-light radius. From the slope of the relation one finds that $r_{\rm half}$ follows $(-E_{\rm tot})^{-1.0}$, with a correlation of $\rho$= $-$0.69 associated to an extremely low probability that it occurs by chance, $P$= 1.2 $\times$ $10^{-23}$. This just indicates that bulge {\rm  GCs} (average $r_{\rm half}$= 2.5 pc) are smaller than {\rm GSE} {\rm  GCs} (average $r_{\rm half}$= 3.6 pc), which are smaller than Sgr {\rm  GCs} (average $r_{\rm half}$= 9.7 pc). Since the total energy gives a proxy of the look-back time (see Figures~\ref{fig:E-h-ecc} and ~\ref{fig:tvsE}), it implies that the half-light radius of {\rm  GCs} is decreasing with the time spent into the MW halo. 

The correlation between half-light radius and total energy followed by the {\rm  GCs} is not similar for dwarfs, which are far much more scattered in the  ($E_{\rm tot}$, $r_{\rm half}$) plane. In particular, many  dwarfs (and all the classical dwarfs) lie well above the correlation. However, the bottom panel of Figure~\ref{fig:E-rh} illustrates a different behavior for {\rm  GCs} and dwarfs with eccentric orbits (ecc$>$ 0.6). First, eccentric {\rm  GCs} show a more robust correlation between their half-light radii and their total energies (77 GCs, $\rho$= $-$0.82, $P$= 1.7 $\times$ $10^{-29}$) also with a slope of $-1.0$. Second, dwarfs appear to fall well within this correlation, thought with a larger scatter. It suggests that all eccentric stellar systems in the MW halo follow the same ($E_{\rm tot}$, $r_{\rm half}$) relation. The average half-light radii of eccentric {\rm  GCs} associated to the Bulge, to {\rm GSE}, to Sgr, and to dwarfs are 1.95, 3.6, 9.8, and 83.5 pc, respectively, implying a decrease with the time elapsed into the halo. 
Such a decrease has been theoretically investigated for {\rm  GCs} \citep{Aguilar1988,Gnedin1999} since eccentric systems passing near their pericenters are likely affected by MW tidal shocks. This increases the internal energy of their stars, leading the least bound to be expelled. Then, after {\rm  GCs} have lost mass, they contract adiabatically when leaving the pericenter towards the apocenter.

This in turn enables the formation of tidal structures formed from stars tidally ejected from the cluster. This is well illustrated by the occurrence of tidal tails in {\rm  GCs} from \citet[see their Table 3]{Zhang2022}: only 9\% (7 among 79) of ecc$<$ 0.6 {\rm  GCs} possess tidal tails, which contrasts with 22\% (17 among 77) of eccentric {\rm  GCs} (ecc $ >$ 0.6). Appendix~\ref{sec:tides} shows that 3 GCs (Pal 4, Pal 14, and Pal 15) are tidally disrupted since they have their tidal radii smaller than $r_{\rm half}$. Two of them possess strong tidal tails: Pal 14 \citep{Sollima2011}, and Pal 15 \citep{Myeong2017}.  Despite the observational efforts \citep{Ibata2021,Zhang2022}, there is still a possibility that some tails have not been discovered yet, and  we would like to suggest deep observations of Pal 4. 

 The most tidally disrupted {\rm  GCs} by the MW that are labelled in Figure~\ref{fig:rh_rp} are {\rm  LSB-GCs}. LSB-GC have sufficiently small densities to be strongly affected by MW tides, despite their large pericenters (average $R_{\rm peri}$: 13.8 {\rm kpc}). The latter value is 6 times larger than that for {\rm  HSB-GCs}, but this is largely compensated by the 300 times lower densities of  LSB-GC compared to {\rm  HSB-GCs}. LSB-{\rm  GCs} have been suspected by \citet{Marchi-Lasch2019} to be recent comers into the MW halo, and we confirm this (see Figures~\ref{fig:E-h-ecc} and \ref{fig:E-rh}), simply because their average total energy ($E_{\rm tot}$= $-$7.9 $\times$ $10^{4}$ $\rm km^{2} s^{-2}$) is significantly higher than that of {\rm  HSB-GCs} ($E_{\rm tot}$= $-$15.7 $\times$ $10^{4}$ $\rm km^{2} s^{-2}$). 

\subsection{Dichotomy between eccentric and non-eccentric dwarfs} 


Almost all {\rm  VPOS}  dwarfs  have low eccentricity, while most other dwarfs (including the 5 LMC related dwarfs) have eccentric orbits. The latter are almost all ultra-faint dwarfs, except Sextans and Leo I\footnote{Leo I appears to be quite exceptional with respect to other classical dwarfs, first from its large distance, second from the fact that it would be the only  unbound classical dwarf for our MW mass model;. this is also true after adopting PM values from the HST  \citep{Casetti-Dinescu2021} since they are very similar to those from Gaia EDR3.}. Bottom panel of Figure~\ref{fig:E-rh} shows that they follow the ($r_{\rm half}$, $E_{\rm tot}$) correlation made by eccentric {\rm  GCs}, which suggests that they could be affected by MW tides as well, as it was proposed by \citet{Hammer2019,Hammer2020}. This is also supported by the tidally disrupted properties of Tucana III \citep{Drlica-Wagner2015}, and perhaps of Triangulum II \citep{Martin2016}. This may further apply to Hercules, which could be also tidally disrupted \citep{Kupper2017,Garling2018}. In the near future, we intend to verify whether all eccentric and energetic ultra-faint dwarfs have morphologies and kinematics consistent with being tidally disrupted.

The dichotomy between low eccentric {\rm  VPOS} dwarfs and other dwarfs with high eccentricity, including those apparently linked to the {\rm LMC} is also reflected in Figure~\ref{fig:E-h-ecc}. {\rm  VPOS}  dwarfs lie near the $V_{\rm rad}$=0 line in Figure~\ref{fig:E-h-ecc}, while almost all other dwarfs have still a significant radial velocity and lie much further from this line. It suggests that dwarfs could have been shaped by different combinations of tides and ram pressure effects:
\begin{itemize}
\item VPOS dwarfs with low eccentricities include almost all classical dwarfs (but Sextans and Leo I), and one may conjecture that it takes more time for ram-pressure  to remove gas from more massive dwarfs; if correct, it means that ram pressure is particularly efficient to slow them down especially along their radial motion, explaining the observed velocity asymmetry for dwarfs \citep{Cautun2017,Hammer2021}, as it comes from especially from massive dwarfs of the {\rm  VPOS};
\item Non-VPOS (and LMC-associated) dwarfs are mostly eccentric and ultra-faint dwarfs, and their gas could have been almost instantaneously removed by ram pressure, which would have preserve their orbit eccentric, while their half-light radii could be affected by both tides and ram pressure;
\item Because of their relatively recent infall, both VPOS and non-VPOS
  dwarfs may have lost their gas  at relatively recent epochs ($<$3 Gyr), leading to an expansion of their residual stars due to a loss of gravity \citep{Hammer2019,Hammer2020}. Such a mechanism could explain why  their half-light radii is much larger than what is expected from the ($r_{\rm half}$, $R_{\rm peri}$) correlation in Figure~\ref{fig:rh_rp}.
\end{itemize}

The above confirms that Gaia EDR3 results are sufficiently robust to avoid specific corrections to dwarf tangential velocities as proposed by \citet{CorreaMagnus2022} and \citet*{Pace2022}, even if this could be necessary for a few extremely faint dwarfs having proper motions with very large error bars. 

\subsection{Some lessons about the MW mass assembly}
Figure~\ref{fig:E-h-ecc} provides useful informations about the origin of each merger having affected the MW. {\rm Pontus}, and {\rm GSE} {\rm  GCs} are all with high eccentricity, while bulge and disk {\rm  GCs} are mostly made of systems with low eccentricities. This indicates that {\rm GSE} and {\rm Pontus} were likely related to a very eccentric infall of their progenitors. Part of the eccentricity of the {\rm GSE} GC orbits may come from an orbit radialization \citep*{Vasiliev2022}, but this could apply only to the {\rm GSE} progenitor, and not to the associated GCs, because \citet{Vasiliev2022} found that radialization is not efficient for very large mass ratios, i.e., GC masses are tiny in comparison to that of the MW or of its bulge. Interestingly, the eccentric {\rm GSE} {\rm  GCs} appear to be more compact than expected from the relation shown in the top panel of Figure~\ref{fig:E-rh}, which is consistent with the fact that they are actively tidally shocked.   

At increasing energies, one finds that the Sgr system eccentricity is in between that of bulge and {\rm GSE} {\rm  GCs} while the Sgr eccentricity is 0.66. However, the situation of the Sgr dwarf appears to be rather exceptional: it is very offset from the ($r_{\rm half}$, $E_{\rm tot}$) correlation of the bottom panel in Figure~\ref{fig:E-rh}, while it possesses the most prominent system of tidal tails. Perhaps this is just due to the fact that, despite its large eccentric orbit, the initial Sgr was so massive that it took more time for the MW to strip it efficiently and have it joining the correlation of  Figure~\ref{fig:E-rh}. The simulations  by \citet{Wang2022} show that after 1.5-2 Gyr from now, almost all the stars in the extended Sgr will be removed, letting only the nuclear GC NGC 6715 that lie on to the correlation.
One can also wonder how a dwarf like Triangulum II can share the same location in the ($E_{\rm tot}$, $r_{\rm half}$)  plane as Eridanus, while the luminosity of the latter is 40 times larger. The same applies to Tucana III and Pal 15, which both display tidal tails. If both type of stellar systems are recent comers, this just means that MW tides may have affected similarly systems with different masses, and indeed tidal theory shows little effect on host to satellite mass ratio.

\section{Conclusions}

This paper shows that an intrinsic structural parameter such as the half-light radius depends on orbital parameters such as pericenter, orbital eccentricity, angular momentum, and total orbital energy. Orbital parameters have been derived from Gaia EDR3 with an unprecedented accuracy for both GCs (this paper, Appendix~\ref{sec:tables}) and dwarfs \citep{Li2021}.  
It results that the intrinsic GC sizes tightly depend on the orbital parameters, the smaller they are, the smaller their pericenter, angular momentum and orbital energy. We have shown that this is likely due to the impact of MW tidal shocks that progressively shrink GC half-light radii.\\

We have also used the archaeological studies of the former mergers that occurred into the MW \citep{Malhan2022,Kruijssen2020}. Here, we have been able to characterize the infall time of each event from its total orbital energy. We have shown that the latter can be used as a proxy of the epoch when stellar systems have emerged in the MW halo.  This provides a relatively simple interpretation of the properties of MW halo inhabitants:

\begin{itemize}
\item Bulge, {\rm Kraken}, {\rm GSE}, and Sgr {\rm  GCs} are inhabitants of the MW halo since $\sim$ 12-13, 11-12, 8-10, and 4-6 Gyr ago, respectively;
\item Most dwarfs are newcomers into the MW halo less than 3 Gyr ago, which is confirmed by their large total energy and angular momentum, as well as by the large scatter of their distribution in most fundamental relations; 
\item  {\rm  VPOS}  dwarfs that include the most massive ones have a different behavior than other dwarfs in both ($E_{\rm tot}$, h) and ($r_{\rm half}$, $E_{\rm tot}$) planes; first they have a minimal energy for their angular momentum meaning small eccentricity orbits, and second, most of their half-light radii are larger than expectations from the strong correlation between $r_{\rm half}$ and $E_{\rm tot}$ established for GCs, all properties consistent with a recent lost of their gaseous content due to ram-pressure from the MW halo;
\item Many ultra-faint dwarfs unrelated to the {\rm  VPOS}  or related to the LMC possess high orbital energy, suggesting that they came very late in the MW halo. Since their orbits are highly eccentric, and that they share the same dependency between $r_{\rm half}$ and $E_{\rm tot}$ than eccentric GCs,  they could be affected as well by strong tidal effects. Such a behaviour is not unexpected, because even if late arrivals do not favor strong tidal effects, this could be compensated by their very low surface or volume densities.
\end{itemize}

This paper is  the first one to investigate in detail the relation between structural and orbital parameters of MW inhabitants since the pioneering work of  \citet{vandenBergh2011}. It also consolidates all the work made by Gaia EDR3 in predicting robust proper motions and tangential velocities, including their error bars. This is true also for most dwarf galaxies, because otherwise we would have not observed the dichotomy presented by {\rm  VPOS}  and non-{\rm  VPOS}   dwarf properties.\\

The correlations shown in this paper are made using a MW mass model following \citet{Eilers2019}, assuming a total mass of 8.2 $\times$ $10^{11}$ $\rm M_{\odot}$, which has been purposely made for reproducing the  MW rotation curve with a NFW halo. An advantage of this choice, is that all GCs and almost all dwarfs have negative total orbital energies. One may wonder whether the results depend on this choice. We have investigated the 3 other models of the MW made by \citet{Jiao2021} that cover the largest possible mass range (from 2.3 to 15 $\times$ $10^{11}$ $\rm M_{\odot}$) able to reproduce the MW rotation curve. Main results of this paper are unaffected, because they depend on the relative positions of the stellar systems along  the total energy axis, the latter being a proxy of the infall time. \\

We conclude that the MW tides and ram-pressure appear to be quite preponderant in shaping the structure of the halo inhabitants, which opens the road for studying their morphologies and their kinematics, to verify whether these properties can be influenced as well by the host, our Galaxy. This work may help the understanding of GC and dwarf origins, as well as testing sophisticated physical models of MW GCs such as that of \citet{Reina-Campos2022}. It would be interesting to verify whether they could reproduce the ($r_{\rm half}$ and $R_{\rm peri}$) correlation found in Figure~\ref{fig:rh_rp}. In principle, if accounting for tidal and ram-pressure effects due to the MW, the intrinsic dynamical evolution of the stellar systems can be traced back with a good accuracy in time, based on their present-day orbital energy.

\section*{Acknowledgments}

 We are very grateful to Frederic Arenou and Carine Babusiaux for their advices, namely to use GaiaEDR3 data and their statistics. We warmly thank Holger Baumgardt for his help in using his remarkable work in gathering and producing very advanced data for globular clusters. We thank the referee for the very useful remarks that have substantially improved the manuscript.
 We are grateful for the support of the International Research Program Tianguan, which is an agreement between the CNRS in France, NAOC, IHEP, and the Yunnan Univ. in China.
 Marcel S. Pawlowski acknowledges funding of a Leibniz-Junior Research Group (project number J94/2020) and a KT Boost Fund by the German Scholars Organization and Klaus Tschira Stiftung.

\section*{Data Availability}

All necessary data used in this paper are available in the Tables of the Appendix~\ref{sec:tables} and in \citet{Li2021}.



\bibliographystyle{mnras}
\bibliography{references_wAbbrevs} 




\appendix

\section{Relating concentration parameter and surface brightness}
\label{sec:cSB}
   \begin{figure*}
  \centering
\includegraphics[scale=0.28]{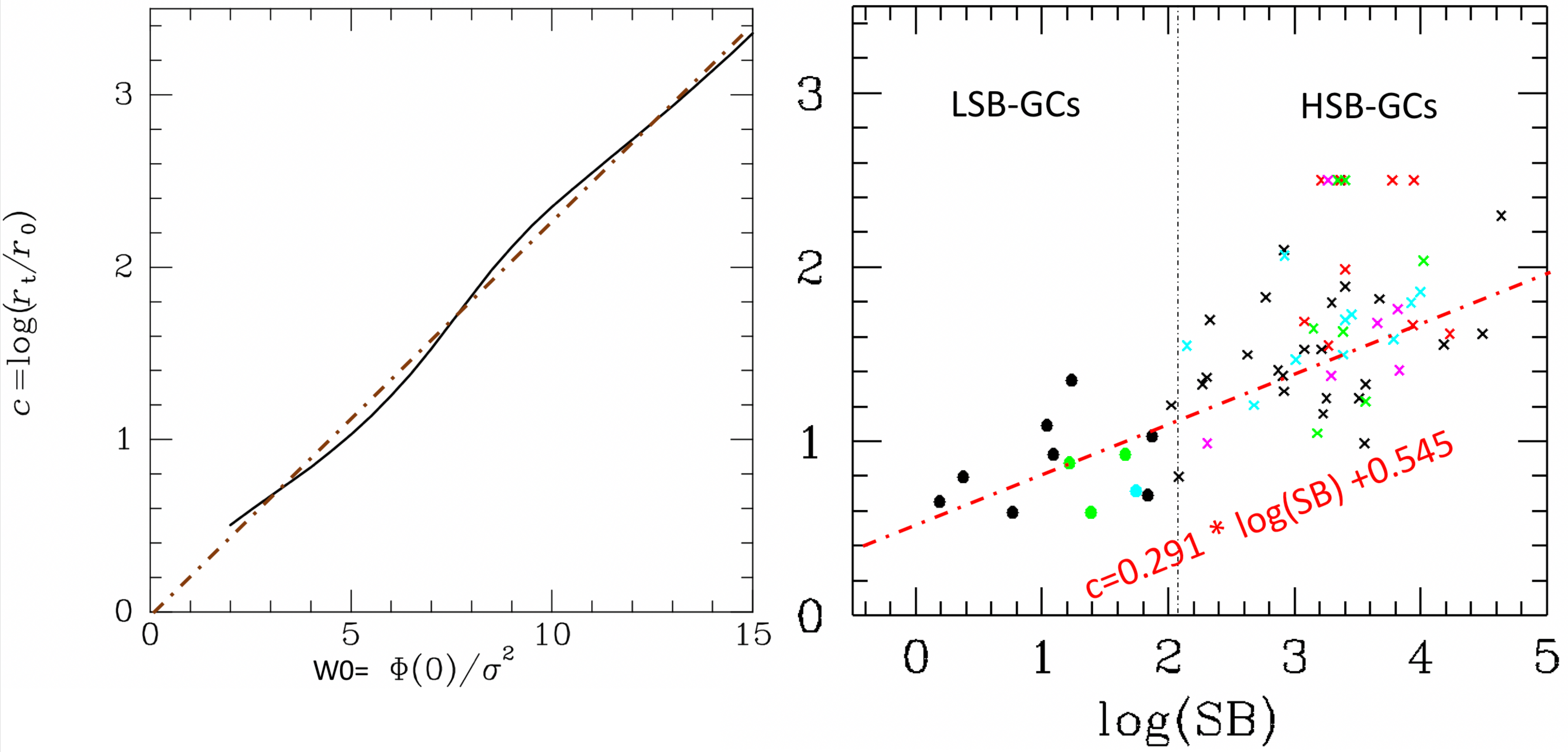}
  \caption{The left panel shows the relationship between King profile compactness c and central concentration parameter $W_{\rm 0}$ based on Figure 4.9 of \citet{Binney2008}. The right panel shows the correlation between the logarithm of the averaged surface brightness inside $r_{\rm half}$ and the King profile compactness c.}
\label{fig:SB_comp}
   \end{figure*}
 
A considerable effort has been done by \citet{Baumgardt2020}, who significantly improved the determination of absolute luminosity, providing corrections that can be as high as two magnitudes in $V$-band, especially for GCs affected by extinction. This implies that the surface brightness calculated by \citet{Baumgardt2020} is the best indicator, currently available, of the GC compactness, and does not depend on extinction effects.  Figure~\ref{fig:SB_comp} allows to convert surface brightness into central concentration parameter $W_{\rm 0}$, after accounting for the correlation between the King compactness $c$ and $\log$(SB) ($\rho$= 0.62 for 70 GCs from \citealt{Harris2010} with eccentricity larger than 0.6). It results that we can approximate:

\begin{equation}
W_0= 4.412\times c \ ,
\end{equation}
and
\begin{equation}
c= 0.291\times \rm \log(SB) +0.545 \ ,
\end{equation}
which leads to:
\begin{equation}
W_0= 1.2834\times \rm \log(SB)+ 2.406 \ .
\label{eq:compactness}
\end{equation}

In the following we adopt Equation~\ref{eq:compactness} for attributing compactness to each individual GCs, and furthermore, assumes that each cluster with $W_{\rm 0}$ $\ge$ 8 ($\le$2) would have its half-light radius shrunk (expanded) by 3\% (by 7\%) at each pericenter passage (see, e.g., \citealt{Martinez-Medina2022}).  Furthermore, we have interpolated the tidal shock effects on GCs for cluster having $W_{\rm 0}$ values from 2 to 8 adopting:
\begin{equation}
\Delta r_{\rm half}/r_{\rm half} = 0.015\times \left(6 - W_0\right) \ .
\end{equation}

Figure~\ref{fig:rh_evolution} illustrates how eccentric GCs are affected by tides, following the scheme from \citet{Martinez-Medina2022}, and deriving the number of pericenter passages from the orbital parameters given in Appendix C.

\section{Tidal radius and search for GCs presently dominated by tidal effects}
\label{sec:tides}
We introduce the tidal radius, $r_{\rm tidal}$, and for quasi circular orbits, we adopt the definition from \citet*{Innanen1983}:
\begin{equation} 
r_{\rm tidal}= \frac{2 R_{\rm peri}}{3}    \left[\frac{m_{\rm GC}}{M_{\rm MW}(R_{\rm peri})}\right]^{1/3}  \left[1-\ln(1 - \rm ecc)\right]^{1/3},
\label{Eq1}
\end{equation} 
in which the last term accounts for the effects related to eccentricity\footnote{Note that the  eccentricity of Eq.~(\ref{Eq1}) differs from the definition that we adopted in Eq.~(\ref{ecc}).}, which we will call later, $\lambda$=$[1-\ln(1-\rm ecc)]^{1/3}$. This term is introduced to compare an eccentric orbit with a circular one with the same total energy. Eq.~\ref{Eq1} further assumes that  the MW cumulative mass increases following the radius ($M_{\rm MW}(R_{\rm peri})$ $\sim$ $R_{\rm peri}$). Such a condition is certainly fulfilled for radii $<$ 20 {\rm kpc}, as shown by \citet{Jiao2021} from their analysis of the MW rotation curve. 

After converting mass into density ($\rho(r)$) and radius ($r$) for both GC and the MW:
 \begin{equation}
 m_{\rm GC}= 2 \, \frac{4}{3}\, \pi\, \rho_{\rm GC}(r_{\rm half}) \, r_{\rm
   half}^{3}  \ ,
\end{equation}
 \begin{equation}
 M_{\rm MW}(R_{\rm peri}) =\frac{4}{3} \,\pi \,\rho_{\rm MW}(R_{\rm peri})\,
 R_{\rm peri}^{3} \ . 
\end{equation}
Eq.~\ref{Eq1} is equivalent to
\begin{equation} 
r_{\rm tidal} \simeq 0.84 \, r_{\rm half} \left[\frac{\rho_{GC}(r_{\rm
      half})}{\rho_{{\rm MW}}(R_{\rm peri})}\right]^{1/3}  \lambda \ ,
\label{Eq2}
\end{equation} 
where $\rho_{GC}(r_{\rm half})$ and $\rho_{\rm MW}(R_{\rm peri})$ are the mean mass densities of the GC inside its half-light radius, and of the MW inside $R_{\rm peri}$, respectively. Eq.~\ref{Eq2} recovers the well-known fact that the tidal radius is that where the mean density matches that of the perturber (the MW) at that position. If $\rho_{GC}(r_{\rm half})$ = $\rho_{\rm MW}(R_{\rm peri})$, Eq.~\ref{Eq2} recovers the fact that the tidal radius matches precisely  the half-mass radius for a GC median eccentricity of 0.6. In such a case, GCs can be considered as being dominated by MW tides. 

However, the tidal formula of \cite{Innanen1983}, as those of \citet{King62}
and \citealt{Bertin&Varri08}, add a centrifugal term, which is only
meaningful if the GCs where phase locked with the Milky Way (they are
insufficiently rigid to be phase locked). Furthermore,
their formulae are only relevant for quasi-circular orbits where the tidal
effects are long-lived.
For  eccentric orbits, the instantaneous tide is short lived and
changes direction, and one should instead treat the tide as a shock and orbit
average it. Following \citet{Mamon00} and its refinement by
\citet{Tollet+17},  a star of a MW satellite should suffer a velocity impulse
equal to the  tidal acceleration at pericenter times the duration $R_{\rm
  peri}/V_{\rm peri}$ of the pericentric passage.  \citet{Mamon00} showed
that that eccentric orbits suffer tides that limit the system to a radius
where the mean density is not the mean density of the host at pericentre, but
lower by roughly the square of the ratio of pericentric to circular velocity
at pericenter. Therefore, at fixed pericenter, the more eccentric orbits
(with higher pericentric velocities) suffer less from tides. This is contrary
to the instantaneous formulae of \cite{King62}, \cite{Innanen1983}, and \cite{Bertin&Varri08}, which
predict, at given pericenter, a  too small tidal radius for eccentric
orbits, and are thus only applicable for quasi-circular orbits and not for
eccentric orbits.


The few {\rm  GCs} (especially Pal 14, Pal 15 and Pal 4) that escape the correlation shown in Figure~\ref{fig:rh_rp} have average density inside $r_{\rm half}$ close to that of the MW at their pericenters, even after accounting for the corrective factor introduced by \citet{Mamon00}. It may indicate that low density {\rm  GCs} on high eccentric {\rm  GCs} are actively shaped by the MW tides, which perturb them sufficiently to move them off the correlation between ($r_{\rm half}$, $R_{\rm peri}$) shown in Figure~\ref{fig:rh_rp}.

Because for all GCs, the half-light radius decreases with the pericenter following a tight 1/3 power law relation, it further suggests that for all GCs, their radii could have been settled by MW tides.

\section{Orbital parameters of 156 GCs}
\label{sec:tables}
\onecolumn

Description of the Table~\ref{tab:sph} content: Column 1 lists the globular cluster name; Column 2-4 is the Galactocentric distance, angle with respect to the North Galactic Pole and azimuthal angle; Column 5-7 gives the velocities in three dimensions; Column 8 provides the Galactocentric tangential velocity and Column 9 lists the total velocity in the Galactic rest frame.\\

Description of the Table~\ref{tab:dyn} content: Column 1 lists the globular cluster name; Column 2 and 3 give the pericenter and the apocenter of the orbit; Column 4 is the eccentricity of the orbit; in Column 5, we provide the probability of the object being unbound. Column 6 gives the orbital phase.

\centering
\begin{longtable}[c]{lcccccccc}
\label{tab:sph}\\
\caption{Kinematic properties of globular clusters in Galactocentric spherical coordinate system.}\\
\hline
\hline
name & $R_\mathrm{GC}$ & $\theta$ & $\phi$ & $V_r$ & $V_\theta$ & $V_\phi$ & $V_\mathrm{tan}$ & $V_\mathrm{3D}$\\
	 & (kpc) & (deg) & (deg) & (km s$^{-1}$) & (km s$^{-1}$) & (km s$^{-1}$) & (km s$^{-1}$) & (km s$^{-1}$)\\
\hline
\vspace{-8pt}
\endhead
\hline
\endfoot
AM 4&$24.6^{+0.7}_{-0.7}$&$49.3^{+0.2}_{-0.2}$&$124.2^{+0.5}_{-0.5}$&$-10.7^{+15.2}_{-16.4}$&$221.6^{+51.1}_{-46.9}$&$92.2^{+50.7}_{-48.9}$&$242.8^{+58.9}_{-55.7}$&$243.0^{+60.1}_{-55.8}$\\
Arp 2&$21.5^{+0.3}_{-0.3}$&$118.3^{+0.1}_{-0.1}$&$192.2^{+0.1}_{-0.0}$&$124.3^{+0.5}_{-0.5}$&$-269.3^{+4.7}_{-4.6}$&$-77.2^{+4.2}_{-4.6}$&$280.1^{+5.3}_{-4.9}$&$306.5^{+4.6}_{-4.3}$\\
BH 140&$6.8^{+0.0}_{-0.0}$&$92.8^{+0.2}_{-0.1}$&$36.3^{+2.2}_{-2.0}$&$192.3^{+13.8}_{-12.8}$&$-41.0^{+2.6}_{-3.0}$&$-114.0^{+7.3}_{-8.8}$&$121.1^{+9.4}_{-7.7}$&$227.2^{+16.8}_{-14.8}$\\
BH 176&$12.7^{+0.9}_{-0.8}$&$83.4^{+0.1}_{-0.1}$&$128.7^{+1.4}_{-1.5}$&$-34.9^{+10.1}_{-8.7}$&$-3.8^{+2.1}_{-2.0}$&$-267.0^{+19.8}_{-22.4}$&$267.0^{+22.4}_{-19.7}$&$269.4^{+23.2}_{-19.5}$\\
BH 184&$4.3^{+0.0}_{-0.0}$&$94.8^{+0.1}_{-0.1}$&$71.5^{+2.1}_{-2.1}$&$57.9^{+1.8}_{-1.8}$&$84.1^{+2.1}_{-1.9}$&$-116.6^{+3.2}_{-3.7}$&$143.9^{+4.0}_{-3.7}$&$155.1^{+4.3}_{-4.1}$\\
BH 229&$1.2^{+0.1}_{-0.1}$&$76.6^{+1.6}_{-1.8}$&$15.3^{+2.2}_{-1.9}$&$-103.2^{+5.8}_{-6.7}$&$226.5^{+2.4}_{-2.8}$&$8.7^{+6.2}_{-6.4}$&$226.7^{+2.7}_{-2.9}$&$249.1^{+5.1}_{-5.2}$\\
BH 261&$2.1^{+0.2}_{-0.2}$&$104.8^{+2.1}_{-1.8}$&$349.9^{+1.3}_{-1.5}$&$110.5^{+18.4}_{-18.9}$&$103.7^{+5.1}_{-4.9}$&$-182.2^{+6.2}_{-5.5}$&$209.9^{+6.9}_{-7.9}$&$237.2^{+2.7}_{-2.0}$\\
Crater&$146.9^{+3.1}_{-3.0}$&$42.0^{+0.0}_{-0.0}$&$90.1^{+0.1}_{-0.1}$&$-7.5^{+2.9}_{-3.9}$&$-81.9^{+63.4}_{-66.3}$&$13.0^{+75.1}_{-50.9}$&$107.9^{+72.4}_{-67.1}$&$108.1^{+72.2}_{-66.7}$\\
Djorg 1&$1.2^{+1.0}_{-0.4}$&$107.5^{+7.2}_{-6.6}$&$153.1^{+11.1}_{-14.7}$&$-250.9^{+78.4}_{-49.5}$&$65.9^{+8.7}_{-16.4}$&$-314.8^{+14.0}_{-34.1}$&$321.5^{+34.8}_{-14.9}$&$406.7^{+23.4}_{-10.0}$\\
Djorg 2&$0.8^{+0.1}_{-0.2}$&$116.9^{+8.8}_{-4.0}$&$216.7^{+21.5}_{-6.9}$&$7.5^{+72.1}_{-27.4}$&$75.8^{+11.4}_{-49.1}$&$192.4^{+3.7}_{-8.0}$&$206.6^{+2.7}_{-12.1}$&$209.0^{+1.5}_{-1.7}$\\
E 1&$120.2^{+2.4}_{-2.3}$&$137.7^{+0.0}_{-0.0}$&$72.7^{+0.1}_{-0.1}$&$-45.0^{+10.6}_{-10.2}$&$-5.4^{+32.8}_{-52.7}$&$7.5^{+33.9}_{-24.7}$&$41.1^{+44.3}_{-27.9}$&$63.0^{+34.4}_{-17.7}$\\
E 3&$9.1^{+0.1}_{-0.1}$&$106.3^{+0.3}_{-0.3}$&$52.6^{+1.3}_{-1.3}$&$25.6^{+1.0}_{-1.1}$&$-109.0^{+3.8}_{-3.7}$&$-234.5^{+4.4}_{-4.7}$&$258.7^{+5.6}_{-5.7}$&$259.9^{+5.7}_{-5.7}$\\
ESO 280-06&$13.5^{+0.6}_{-0.5}$&$109.7^{+0.3}_{-0.3}$&$158.5^{+0.4}_{-0.4}$&$61.0^{+1.0}_{-1.0}$&$55.5^{+3.8}_{-3.6}$&$-36.3^{+8.4}_{-8.9}$&$66.4^{+7.5}_{-6.4}$&$90.2^{+6.4}_{-5.4}$\\
ESO 452-11&$2.1^{+0.1}_{-0.0}$&$43.1^{+3.0}_{-3.2}$&$43.7^{+6.9}_{-5.2}$&$-94.5^{+2.4}_{-2.0}$&$37.4^{+1.4}_{-1.2}$&$-10.8^{+8.0}_{-7.4}$&$38.9^{+3.7}_{-2.3}$&$102.2^{+1.1}_{-0.9}$\\
ESO 93-8&$12.8^{+0.5}_{-0.5}$&$94.2^{+0.0}_{-0.0}$&$77.9^{+1.5}_{-1.6}$&$-4.1^{+6.2}_{-6.7}$&$6.1^{+1.4}_{-1.5}$&$-218.1^{+12.2}_{-11.7}$&$218.2^{+11.8}_{-12.1}$&$218.3^{+11.6}_{-12.3}$\\
Eridanus&$89.7^{+2.1}_{-2.1}$&$128.5^{+0.1}_{-0.1}$&$34.0^{+0.1}_{-0.1}$&$-141.5^{+1.3}_{-1.1}$&$-65.7^{+16.2}_{-16.0}$&$-8.9^{+12.1}_{-13.8}$&$67.9^{+15.6}_{-16.4}$&$156.7^{+6.7}_{-5.3}$\\
FSR 1716&$4.1^{+0.0}_{-0.0}$&$92.5^{+0.1}_{-0.1}$&$65.6^{+3.5}_{-3.6}$&$101.5^{+1.8}_{-2.2}$&$111.8^{+4.4}_{-4.2}$&$-157.0^{+6.9}_{-7.8}$&$192.7^{+8.9}_{-7.9}$&$217.7^{+8.8}_{-7.9}$\\
FSR 1735&$3.2^{+0.2}_{-0.2}$&$94.7^{+0.0}_{-0.1}$&$94.3^{+7.2}_{-9.5}$&$-121.6^{+8.8}_{-5.3}$&$-103.7^{+5.6}_{-5.1}$&$-104.8^{+12.3}_{-16.4}$&$147.8^{+8.4}_{-5.1}$&$191.9^{+3.0}_{-3.3}$\\
FSR 1758&$3.2^{+0.8}_{-0.8}$&$100.6^{+2.1}_{-1.4}$&$140.5^{+6.3}_{-10.6}$&$15.8^{+36.9}_{-63.4}$&$-193.0^{+28.9}_{-21.2}$&$328.6^{+2.6}_{-4.9}$&$382.2^{+7.0}_{-12.6}$&$382.5^{+10.2}_{-10.1}$\\
IC 1257&$19.2^{+1.1}_{-1.6}$&$68.8^{+0.3}_{-0.5}$&$204.0^{+0.7}_{-0.4}$&$-59.8^{+2.4}_{-2.1}$&$-4.8^{+3.3}_{-3.7}$&$9.8^{+15.6}_{-9.9}$&$12.9^{+13.3}_{-7.5}$&$62.1^{+2.3}_{-1.9}$\\
IC 4499&$15.7^{+0.2}_{-0.2}$&$114.7^{+0.0}_{-0.0}$&$100.5^{+0.3}_{-0.3}$&$-180.1^{+0.8}_{-0.8}$&$150.5^{+2.2}_{-2.1}$&$72.9^{+2.1}_{-2.1}$&$167.3^{+2.3}_{-2.4}$&$245.9^{+1.8}_{-1.8}$\\
Laevens 3&$58.9^{+1.2}_{-1.2}$&$112.3^{+0.0}_{-0.0}$&$251.3^{+0.2}_{-0.2}$&$129.3^{+2.4}_{-2.6}$&$72.2^{+27.4}_{-26.0}$&$-36.6^{+18.0}_{-19.6}$&$83.7^{+25.3}_{-23.9}$&$154.4^{+14.9}_{-11.9}$\\
Liller 1&$0.7^{+0.0}_{-0.0}$&$89.8^{+0.1}_{-0.0}$&$82.6^{+17.3}_{-15.2}$&$108.8^{+20.1}_{-25.6}$&$-24.9^{+2.3}_{-2.5}$&$52.5^{+24.2}_{-37.3}$&$58.7^{+21.7}_{-23.6}$&$123.7^{+8.9}_{-8.2}$\\
Mercer 5&$3.2^{+0.3}_{-0.3}$&$89.8^{+0.0}_{-0.0}$&$328.3^{+5.6}_{-7.3}$&$-169.8^{+11.5}_{-7.1}$&$-10.5^{+1.8}_{-1.9}$&$-207.4^{+6.4}_{-9.9}$&$207.6^{+10.2}_{-6.4}$&$268.4^{+2.9}_{-2.8}$\\
Munoz 1&$47.3^{+1.5}_{-1.5}$&$47.1^{+0.1}_{-0.1}$&$298.5^{+0.4}_{-0.4}$&$21.2^{+6.3}_{-5.8}$&$118.4^{+30.0}_{-29.3}$&$-89.9^{+38.9}_{-37.6}$&$151.5^{+36.0}_{-38.8}$&$153.0^{+35.7}_{-37.7}$\\
NGC 104&$7.5^{+0.0}_{-0.0}$&$115.1^{+0.1}_{-0.1}$&$22.6^{+0.2}_{-0.1}$&$-8.8^{+0.4}_{-0.3}$&$-46.3^{+0.4}_{-0.4}$&$-176.7^{+0.5}_{-0.5}$&$182.7^{+0.4}_{-0.4}$&$182.9^{+0.3}_{-0.4}$\\
NGC 1261&$18.2^{+0.1}_{-0.1}$&$135.1^{+0.1}_{-0.1}$&$51.4^{+0.3}_{-0.3}$&$-107.8^{+0.8}_{-0.9}$&$11.2^{+1.7}_{-1.8}$&$32.7^{+2.9}_{-2.7}$&$34.6^{+2.4}_{-2.3}$&$113.2^{+1.5}_{-1.4}$\\
NGC 1851&$16.6^{+0.1}_{-0.1}$&$114.3^{+0.1}_{-0.1}$&$35.6^{+0.1}_{-0.2}$&$138.8^{+1.0}_{-1.1}$&$28.7^{+1.8}_{-1.9}$&$16.4^{+1.1}_{-1.1}$&$33.1^{+1.7}_{-1.6}$&$142.6^{+1.1}_{-1.2}$\\
NGC 1904&$19.0^{+0.2}_{-0.1}$&$109.6^{+0.1}_{-0.1}$&$27.8^{+0.1}_{-0.1}$&$48.7^{+0.5}_{-0.5}$&$-22.9^{+2.0}_{-2.1}$&$4.9^{+2.2}_{-2.0}$&$23.4^{+2.5}_{-2.3}$&$54.1^{+0.9}_{-1.0}$\\
NGC 2298&$15.0^{+0.2}_{-0.1}$&$100.3^{+0.1}_{-0.1}$&$35.6^{+0.3}_{-0.3}$&$-87.4^{+1.2}_{-1.3}$&$-52.5^{+1.7}_{-1.9}$&$28.0^{+3.1}_{-3.1}$&$59.4^{+3.0}_{-2.8}$&$105.6^{+2.8}_{-2.4}$\\
NGC 2419&$96.0^{+1.7}_{-1.8}$&$66.8^{+0.0}_{-0.0}$&$0.3^{+0.0}_{-0.0}$&$-27.0^{+0.5}_{-0.4}$&$44.7^{+10.5}_{-10.4}$&$-27.0^{+10.3}_{-9.8}$&$53.5^{+10.7}_{-12.6}$&$60.0^{+9.5}_{-10.7}$\\
NGC 2808&$11.5^{+0.1}_{-0.1}$&$99.7^{+0.0}_{-0.0}$&$57.9^{+0.4}_{-0.3}$&$-148.2^{+0.8}_{-0.8}$&$-5.0^{+0.9}_{-0.9}$&$-26.1^{+1.5}_{-1.5}$&$26.6^{+1.4}_{-1.4}$&$150.6^{+0.6}_{-0.6}$\\
NGC 288&$12.2^{+0.1}_{-0.1}$&$137.5^{+0.3}_{-0.3}$&$359.7^{+0.0}_{-0.0}$&$-33.4^{+0.6}_{-0.5}$&$-37.6^{+0.5}_{-0.6}$&$61.1^{+3.0}_{-3.5}$&$71.7^{+2.6}_{-2.9}$&$79.1^{+2.3}_{-2.7}$\\
NGC 3201&$8.9^{+0.0}_{-0.0}$&$85.2^{+0.0}_{-0.0}$&$31.7^{+0.2}_{-0.2}$&$-88.0^{+0.3}_{-0.4}$&$-156.8^{+0.7}_{-0.8}$&$310.4^{+1.0}_{-1.0}$&$347.8^{+1.2}_{-1.2}$&$358.7^{+1.2}_{-1.2}$\\
NGC 362&$9.6^{+0.1}_{-0.1}$&$131.5^{+0.3}_{-0.3}$&$46.5^{+0.5}_{-0.6}$&$153.9^{+1.9}_{-1.8}$&$-45.2^{+3.7}_{-4.1}$&$7.4^{+1.2}_{-1.3}$&$45.8^{+3.9}_{-3.5}$&$160.6^{+2.8}_{-2.7}$\\
NGC 4147&$20.7^{+0.1}_{-0.2}$&$29.2^{+0.2}_{-0.1}$&$22.8^{+0.1}_{-0.2}$&$137.2^{+1.0}_{-1.0}$&$-20.3^{+2.9}_{-3.0}$&$7.0^{+2.2}_{-2.2}$&$21.6^{+2.7}_{-2.5}$&$139.0^{+0.6}_{-0.6}$\\
NGC 4372&$7.2^{+0.0}_{-0.0}$&$97.7^{+0.2}_{-0.2}$&$42.8^{+1.5}_{-1.5}$&$16.7^{+2.4}_{-2.6}$&$-67.3^{+3.0}_{-2.9}$&$-121.7^{+1.6}_{-1.7}$&$139.1^{+2.8}_{-2.8}$&$140.1^{+3.1}_{-3.1}$\\
NGC 4590&$10.3^{+0.1}_{-0.1}$&$53.5^{+0.1}_{-0.1}$&$61.5^{+0.4}_{-0.3}$&$-120.0^{+0.8}_{-0.7}$&$-109.4^{+1.7}_{-1.6}$&$-283.3^{+0.8}_{-0.9}$&$303.7^{+1.1}_{-1.1}$&$326.5^{+1.0}_{-0.8}$\\
NGC 4833&$7.1^{+0.0}_{-0.0}$&$97.1^{+0.1}_{-0.1}$&$49.5^{+0.6}_{-0.6}$&$117.9^{+2.5}_{-2.8}$&$27.9^{+0.6}_{-0.6}$&$-29.5^{+2.1}_{-2.0}$&$40.7^{+1.6}_{-1.5}$&$124.6^{+2.9}_{-3.0}$\\
NGC 5024&$19.0^{+0.1}_{-0.1}$&$16.5^{+0.2}_{-0.2}$&$16.0^{+0.2}_{-0.2}$&$-96.2^{+0.7}_{-0.6}$&$-68.9^{+1.7}_{-1.5}$&$-122.5^{+2.5}_{-2.4}$&$140.5^{+2.2}_{-2.3}$&$170.3^{+1.9}_{-2.1}$\\
NGC 5053&$18.0^{+0.2}_{-0.2}$&$16.9^{+0.3}_{-0.2}$&$15.3^{+0.2}_{-0.3}$&$8.1^{+0.7}_{-0.7}$&$-93.6^{+1.5}_{-1.4}$&$-120.2^{+2.6}_{-2.7}$&$152.2^{+2.5}_{-2.2}$&$152.5^{+2.4}_{-2.2}$\\
NGC 5139&$6.5^{+0.0}_{-0.0}$&$77.2^{+0.1}_{-0.1}$&$40.2^{+0.4}_{-0.5}$&$-68.1^{+1.1}_{-1.2}$&$74.3^{+1.5}_{-1.7}$&$87.3^{+0.7}_{-0.8}$&$114.6^{+1.5}_{-1.7}$&$133.3^{+0.8}_{-0.8}$\\
NGC 5272&$12.1^{+0.1}_{-0.1}$&$34.1^{+0.2}_{-0.2}$&$348.6^{+0.1}_{-0.1}$&$-133.6^{+0.6}_{-0.7}$&$42.5^{+1.1}_{-1.2}$&$-126.3^{+1.1}_{-1.2}$&$133.2^{+1.4}_{-1.3}$&$188.7^{+0.9}_{-0.8}$\\
NGC 5286&$8.5^{+0.1}_{-0.1}$&$75.9^{+0.0}_{-0.0}$&$83.9^{+0.6}_{-0.6}$&$-195.3^{+0.8}_{-0.8}$&$-57.8^{+1.0}_{-1.1}$&$36.7^{+2.4}_{-2.5}$&$68.5^{+1.7}_{-1.7}$&$207.0^{+0.9}_{-1.0}$\\
NGC 5466&$16.5^{+0.1}_{-0.1}$&$20.0^{+0.2}_{-0.2}$&$327.2^{+0.4}_{-0.4}$&$265.7^{+0.8}_{-1.0}$&$71.8^{+1.3}_{-1.1}$&$157.6^{+5.4}_{-5.3}$&$173.3^{+5.0}_{-4.9}$&$317.2^{+3.3}_{-3.4}$\\
NGC 5634&$21.9^{+0.5}_{-0.5}$&$25.9^{+0.3}_{-0.3}$&$147.2^{+0.5}_{-0.6}$&$-39.6^{+0.9}_{-0.8}$&$-22.0^{+2.5}_{-2.6}$&$-55.0^{+5.9}_{-5.7}$&$59.2^{+5.1}_{-5.1}$&$71.3^{+4.0}_{-3.9}$\\
NGC 5694&$29.2^{+0.5}_{-0.5}$&$52.9^{+0.1}_{-0.1}$&$141.3^{+0.2}_{-0.2}$&$-241.1^{+0.9}_{-0.9}$&$32.6^{+5.4}_{-5.0}$&$32.9^{+4.1}_{-4.1}$&$46.5^{+5.0}_{-4.6}$&$245.6^{+1.8}_{-1.6}$\\
NGC 5824&$25.5^{+0.5}_{-0.6}$&$62.0^{+0.1}_{-0.1}$&$143.0^{+0.2}_{-0.2}$&$-114.4^{+0.8}_{-0.8}$&$143.5^{+5.4}_{-5.4}$&$-118.3^{+6.9}_{-6.6}$&$185.9^{+7.9}_{-8.1}$&$218.4^{+6.4}_{-6.9}$\\
NGC 5897&$7.4^{+0.2}_{-0.2}$&$31.4^{+0.6}_{-0.6}$&$125.1^{+1.9}_{-1.9}$&$127.4^{+1.4}_{-1.3}$&$33.9^{+1.4}_{-1.6}$&$-110.7^{+8.9}_{-9.4}$&$115.7^{+8.8}_{-8.0}$&$172.3^{+6.7}_{-6.3}$\\
NGC 5904&$6.3^{+0.0}_{-0.0}$&$29.0^{+0.5}_{-0.5}$&$353.5^{+0.1}_{-0.1}$&$-299.3^{+1.2}_{-1.1}$&$-165.5^{+3.3}_{-3.4}$&$-111.4^{+0.6}_{-0.7}$&$199.5^{+2.8}_{-2.6}$&$359.7^{+2.3}_{-2.5}$\\
NGC 5927&$4.8^{+0.0}_{-0.0}$&$81.2^{+0.0}_{-0.0}$&$74.6^{+0.9}_{-0.9}$&$-41.8^{+1.7}_{-1.6}$&$-12.3^{+0.8}_{-0.8}$&$-225.8^{+0.4}_{-0.4}$&$226.2^{+0.4}_{-0.4}$&$230.0^{+0.5}_{-0.4}$\\
NGC 5946&$5.3^{+0.3}_{-0.3}$&$82.0^{+0.0}_{-0.0}$&$91.5^{+4.1}_{-4.6}$&$51.2^{+8.8}_{-9.8}$&$-93.4^{+3.0}_{-3.2}$&$-12.2^{+9.1}_{-10.4}$&$94.2^{+4.8}_{-3.7}$&$107.2^{+8.6}_{-7.5}$\\
NGC 5986&$4.9^{+0.1}_{-0.1}$&$59.9^{+0.2}_{-0.2}$&$108.4^{+1.1}_{-1.3}$&$60.4^{+1.8}_{-1.8}$&$50.6^{+1.6}_{-1.7}$&$-36.0^{+4.0}_{-3.7}$&$62.0^{+3.5}_{-3.5}$&$86.5^{+3.7}_{-3.7}$\\
NGC 6093&$4.0^{+0.1}_{-0.1}$&$29.8^{+0.9}_{-0.9}$&$141.2^{+1.6}_{-1.7}$&$-32.9^{+1.0}_{-1.1}$&$70.0^{+1.9}_{-2.0}$&$-42.5^{+4.2}_{-4.0}$&$81.9^{+3.7}_{-3.7}$&$88.2^{+3.3}_{-3.1}$\\
NGC 6101&$10.4^{+0.1}_{-0.1}$&$112.2^{+0.0}_{-0.0}$&$103.1^{+0.4}_{-0.4}$&$69.4^{+2.0}_{-2.0}$&$177.3^{+1.5}_{-1.4}$&$304.0^{+1.5}_{-1.3}$&$352.0^{+1.7}_{-1.6}$&$358.8^{+1.7}_{-1.6}$\\
NGC 6121&$6.4^{+0.0}_{-0.0}$&$85.2^{+0.1}_{-0.1}$&$2.5^{+0.0}_{-0.0}$&$-56.6^{+0.3}_{-0.3}$&$-1.4^{+0.3}_{-0.3}$&$-30.6^{+2.1}_{-2.0}$&$30.7^{+2.0}_{-2.1}$&$64.4^{+1.2}_{-1.2}$\\
NGC 6139&$3.6^{+0.3}_{-0.3}$&$69.7^{+0.7}_{-0.9}$&$115.6^{+4.3}_{-6.1}$&$54.3^{+1.1}_{-1.5}$&$-120.0^{+5.4}_{-5.0}$&$-82.6^{+10.0}_{-9.3}$&$145.6^{+9.5}_{-9.7}$&$155.5^{+9.1}_{-9.8}$\\
NGC 6144&$2.5^{+0.0}_{-0.0}$&$27.4^{+0.7}_{-0.5}$&$72.0^{+4.5}_{-4.8}$&$-28.5^{+6.9}_{-8.0}$&$-150.3^{+11.4}_{-11.2}$&$145.1^{+11.1}_{-13.2}$&$208.7^{+0.8}_{-0.9}$&$210.8^{+1.0}_{-1.0}$\\
NGC 6171&$3.7^{+0.0}_{-0.0}$&$53.1^{+1.1}_{-1.0}$&$354.1^{+0.2}_{-0.2}$&$-35.4^{+1.6}_{-1.5}$&$45.8^{+0.4}_{-0.4}$&$-81.9^{+2.2}_{-2.3}$&$93.8^{+2.1}_{-2.1}$&$100.3^{+1.6}_{-1.5}$\\
NGC 6205&$8.6^{+0.0}_{-0.0}$&$55.5^{+0.2}_{-0.2}$&$317.4^{+0.4}_{-0.4}$&$-40.3^{+1.2}_{-1.1}$&$66.5^{+1.7}_{-1.7}$&$41.6^{+0.6}_{-0.6}$&$78.5^{+1.2}_{-1.2}$&$88.2^{+0.6}_{-0.6}$\\
NGC 6218&$4.5^{+0.0}_{-0.0}$&$59.7^{+0.4}_{-0.4}$&$341.5^{+0.3}_{-0.3}$&$-54.5^{+1.1}_{-1.2}$&$67.3^{+0.3}_{-0.3}$&$-112.4^{+1.0}_{-1.0}$&$131.1^{+0.9}_{-1.0}$&$141.9^{+0.6}_{-0.6}$\\
NGC 6229&$29.5^{+0.4}_{-0.4}$&$48.6^{+0.0}_{-0.0}$&$274.3^{+0.3}_{-0.3}$&$43.5^{+1.0}_{-0.9}$&$-24.3^{+4.9}_{-4.8}$&$-7.9^{+2.6}_{-2.7}$&$25.7^{+4.8}_{-4.9}$&$50.7^{+2.4}_{-2.2}$\\
NGC 6235&$4.5^{+0.3}_{-0.4}$&$51.3^{+1.8}_{-2.3}$&$176.4^{+0.2}_{-0.3}$&$96.0^{+3.0}_{-3.7}$&$133.1^{+1.8}_{-1.5}$&$-230.9^{+16.3}_{-15.4}$&$266.4^{+13.2}_{-13.1}$&$283.2^{+13.3}_{-13.6}$\\
NGC 6254&$4.3^{+0.0}_{-0.0}$&$62.2^{+0.5}_{-0.4}$&$341.4^{+0.4}_{-0.4}$&$-60.2^{+0.5}_{-0.6}$&$-86.5^{+0.8}_{-0.8}$&$-103.5^{+1.1}_{-1.2}$&$134.9^{+0.5}_{-0.5}$&$147.7^{+0.6}_{-0.5}$\\
NGC 6256&$1.9^{+0.1}_{-0.1}$&$76.2^{+1.2}_{-1.0}$&$57.9^{+8.1}_{-7.4}$&$-38.9^{+16.6}_{-16.9}$&$-79.0^{+7.2}_{-8.4}$&$-174.1^{+10.7}_{-7.9}$&$191.3^{+4.3}_{-6.1}$&$195.2^{+1.9}_{-1.9}$\\
NGC 6266&$2.1^{+0.1}_{-0.1}$&$66.6^{+1.1}_{-1.2}$&$21.5^{+1.3}_{-1.2}$&$72.1^{+0.5}_{-0.5}$&$-37.5^{+0.6}_{-0.6}$&$-114.1^{+1.2}_{-1.3}$&$120.1^{+1.3}_{-1.2}$&$140.1^{+1.2}_{-1.1}$\\
NGC 6273&$1.5^{+0.0}_{-0.0}$&$18.7^{+1.8}_{-0.6}$&$103.5^{+15.9}_{-20.0}$&$107.6^{+13.6}_{-16.7}$&$-218.2^{+45.7}_{-46.2}$&$174.2^{+40.6}_{-69.5}$&$279.7^{+4.8}_{-4.5}$&$299.6^{+1.1}_{-1.1}$\\
NGC 6284&$6.4^{+0.4}_{-0.4}$&$67.3^{+0.7}_{-0.8}$&$176.1^{+0.1}_{-0.2}$&$54.3^{+0.7}_{-0.7}$&$-91.3^{+3.8}_{-3.9}$&$3.5^{+6.8}_{-6.0}$&$91.4^{+3.8}_{-3.4}$&$106.3^{+3.2}_{-2.8}$\\
NGC 6287&$1.6^{+0.1}_{-0.0}$&$16.7^{+11.0}_{-9.8}$&$357.8^{+1.0}_{-4.4}$&$142.9^{+44.1}_{-49.1}$&$266.7^{+25.4}_{-38.7}$&$-62.4^{+30.1}_{-11.7}$&$279.6^{+20.7}_{-26.8}$&$314.4^{+1.1}_{-1.1}$\\
NGC 6293&$1.5^{+0.3}_{-0.2}$&$33.6^{+8.8}_{-8.1}$&$153.5^{+7.5}_{-11.9}$&$-211.3^{+8.5}_{-2.4}$&$-50.2^{+44.6}_{-40.6}$&$52.7^{+15.0}_{-31.4}$&$73.9^{+19.7}_{-6.4}$&$223.7^{+1.0}_{-0.8}$\\
NGC 6304&$2.1^{+0.1}_{-0.1}$&$73.7^{+1.1}_{-1.2}$&$12.5^{+1.0}_{-0.9}$&$94.0^{+0.8}_{-0.8}$&$-50.6^{+0.6}_{-0.6}$&$-174.7^{+1.0}_{-0.9}$&$181.9^{+0.9}_{-0.9}$&$204.7^{+1.0}_{-1.1}$\\
NGC 6316&$3.2^{+0.4}_{-0.4}$&$69.1^{+1.6}_{-2.1}$&$169.5^{+0.9}_{-1.2}$&$129.6^{+1.0}_{-1.2}$&$-49.5^{+7.6}_{-6.6}$&$-96.1^{+14.6}_{-14.1}$&$108.2^{+15.3}_{-16.5}$&$168.8^{+9.5}_{-9.5}$\\
NGC 6325&$1.3^{+0.2}_{-0.1}$&$32.1^{+11.5}_{-13.1}$&$349.1^{+3.7}_{-9.3}$&$23.2^{+13.6}_{-12.9}$&$-101.1^{+9.6}_{-25.0}$&$178.6^{+6.2}_{-15.8}$&$208.2^{+15.1}_{-16.8}$&$209.4^{+16.8}_{-17.8}$\\
NGC 6333&$1.8^{+0.0}_{-0.0}$&$26.8^{+0.5}_{-0.1}$&$269.1^{+9.4}_{-9.1}$&$114.0^{+22.0}_{-23.4}$&$79.4^{+41.7}_{-47.0}$&$-326.4^{+23.2}_{-16.1}$&$335.8^{+8.1}_{-9.0}$&$354.8^{+1.8}_{-1.6}$\\
NGC 6341&$9.8^{+0.0}_{-0.0}$&$60.2^{+0.1}_{-0.1}$&$310.6^{+0.3}_{-0.3}$&$80.4^{+0.6}_{-0.6}$&$-67.0^{+1.5}_{-1.7}$&$0.7^{+1.0}_{-0.9}$&$67.0^{+1.7}_{-1.6}$&$104.7^{+1.1}_{-1.1}$\\
NGC 6342&$1.6^{+0.0}_{-0.0}$&$28.2^{+3.6}_{-1.8}$&$293.8^{+14.0}_{-18.8}$&$-70.2^{+17.4}_{-15.1}$&$-68.9^{+33.1}_{-15.1}$&$-130.4^{+22.3}_{-23.8}$&$147.8^{+10.7}_{-10.7}$&$163.7^{+3.3}_{-2.6}$\\
NGC 6352&$3.5^{+0.0}_{-0.0}$&$101.1^{+0.2}_{-0.2}$&$31.1^{+0.7}_{-0.7}$&$45.8^{+1.6}_{-1.7}$&$-14.9^{+0.6}_{-0.5}$&$-213.6^{+1.0}_{-1.0}$&$214.2^{+1.0}_{-1.0}$&$219.0^{+1.2}_{-1.1}$\\
NGC 6355&$0.9^{+0.1}_{-0.1}$&$18.2^{+14.5}_{-9.7}$&$167.0^{+6.3}_{-18.6}$&$56.5^{+39.0}_{-55.3}$&$-250.7^{+3.4}_{-1.4}$&$69.1^{+20.0}_{-71.8}$&$260.1^{+6.7}_{-12.5}$&$266.3^{+1.2}_{-0.9}$\\
NGC 6356&$7.9^{+0.9}_{-0.9}$&$69.3^{+1.1}_{-1.3}$&$194.1^{+1.0}_{-0.8}$&$87.4^{+2.0}_{-1.7}$&$-85.0^{+8.2}_{-7.6}$&$-133.6^{+19.8}_{-20.5}$&$158.4^{+21.8}_{-21.0}$&$181.0^{+18.4}_{-17.7}$\\
NGC 6362&$5.2^{+0.0}_{-0.0}$&$116.3^{+0.2}_{-0.2}$&$63.0^{+0.8}_{-0.8}$&$-18.3^{+0.6}_{-0.6}$&$-102.2^{+1.2}_{-1.2}$&$-118.2^{+0.5}_{-0.5}$&$156.2^{+1.1}_{-1.1}$&$157.3^{+1.1}_{-1.0}$\\
NGC 6366&$5.2^{+0.0}_{-0.0}$&$79.2^{+0.2}_{-0.2}$&$348.2^{+0.3}_{-0.2}$&$79.2^{+0.5}_{-0.5}$&$78.7^{+0.7}_{-0.8}$&$-118.6^{+1.3}_{-1.3}$&$142.4^{+0.7}_{-0.8}$&$162.9^{+0.9}_{-0.9}$\\
NGC 6380&$2.1^{+0.2}_{-0.2}$&$104.8^{+1.2}_{-0.9}$&$128.0^{+4.3}_{-7.2}$&$-66.1^{+4.5}_{-6.6}$&$5.9^{+3.4}_{-2.3}$&$23.7^{+1.9}_{-4.0}$&$24.4^{+1.7}_{-2.8}$&$70.5^{+5.3}_{-3.8}$\\
NGC 6388&$4.0^{+0.1}_{-0.1}$&$108.6^{+0.3}_{-0.3}$&$133.5^{+1.1}_{-1.1}$&$-7.8^{+2.7}_{-2.6}$&$23.1^{+1.3}_{-1.3}$&$80.7^{+1.8}_{-2.0}$&$84.0^{+1.9}_{-2.2}$&$84.4^{+2.2}_{-2.4}$\\
NGC 6397&$6.0^{+0.0}_{-0.0}$&$94.7^{+0.0}_{-0.0}$&$8.7^{+0.1}_{-0.1}$&$54.9^{+0.7}_{-0.7}$&$126.7^{+0.9}_{-0.9}$&$-102.3^{+0.7}_{-0.7}$&$162.8^{+0.3}_{-0.3}$&$171.9^{+0.5}_{-0.5}$\\
NGC 6401&$0.8^{+0.0}_{-0.0}$&$40.4^{+2.9}_{-0.7}$&$268.4^{+18.7}_{-20.0}$&$233.0^{+15.1}_{-28.6}$&$90.8^{+15.5}_{-45.9}$&$109.1^{+68.2}_{-77.0}$&$143.9^{+39.6}_{-29.9}$&$273.9^{+1.3}_{-1.4}$\\
NGC 6402&$4.0^{+0.1}_{-0.1}$&$53.7^{+0.1}_{-0.0}$&$267.6^{+3.6}_{-3.5}$&$-13.9^{+2.4}_{-2.4}$&$-34.6^{+2.7}_{-2.5}$&$-47.2^{+4.2}_{-4.8}$&$58.5^{+5.4}_{-4.9}$&$60.1^{+5.8}_{-5.4}$\\
NGC 6426&$14.5^{+0.3}_{-0.3}$&$66.3^{+0.2}_{-0.2}$&$224.8^{+0.4}_{-0.4}$&$-123.9^{+1.1}_{-1.2}$&$-23.0^{+2.0}_{-1.9}$&$-106.0^{+6.9}_{-7.0}$&$108.5^{+6.9}_{-6.9}$&$164.6^{+5.5}_{-5.2}$\\
NGC 6440&$1.2^{+0.1}_{-0.0}$&$62.9^{+0.6}_{-0.2}$&$269.1^{+11.2}_{-10.9}$&$47.7^{+9.6}_{-10.9}$&$67.4^{+3.0}_{-4.7}$&$35.9^{+12.0}_{-15.1}$&$76.3^{+2.8}_{-3.0}$&$90.1^{+3.3}_{-2.8}$\\
NGC 6441&$4.8^{+0.1}_{-0.2}$&$103.0^{+0.3}_{-0.2}$&$162.3^{+0.3}_{-0.4}$&$26.5^{+0.5}_{-0.5}$&$20.1^{+1.4}_{-1.3}$&$-114.8^{+4.8}_{-4.6}$&$116.5^{+4.7}_{-4.8}$&$119.4^{+4.6}_{-4.7}$\\
NGC 6453&$2.2^{+0.2}_{-0.2}$&$107.7^{+1.2}_{-1.1}$&$158.5^{+1.6}_{-1.7}$&$-64.8^{+2.7}_{-2.3}$&$166.3^{+2.4}_{-2.2}$&$-34.0^{+1.7}_{-2.1}$&$169.7^{+2.8}_{-2.4}$&$181.7^{+3.1}_{-3.1}$\\
NGC 6496&$2.8^{+0.1}_{-0.1}$&$125.8^{+0.7}_{-0.7}$&$120.5^{+2.5}_{-2.8}$&$46.0^{+6.5}_{-5.8}$&$23.9^{+5.8}_{-6.5}$&$-273.9^{+6.0}_{-5.6}$&$274.9^{+6.1}_{-6.6}$&$278.8^{+5.2}_{-5.3}$\\
NGC 6517&$3.2^{+0.2}_{-0.2}$&$70.1^{+0.5}_{-0.3}$&$260.7^{+10.8}_{-6.7}$&$31.6^{+4.6}_{-3.3}$&$45.2^{+1.1}_{-1.1}$&$-41.7^{+1.8}_{-3.9}$&$61.8^{+2.5}_{-1.7}$&$69.1^{+4.1}_{-2.0}$\\
NGC 6522&$0.9^{+0.2}_{-0.1}$&$123.3^{+9.1}_{-6.7}$&$349.6^{+2.5}_{-4.1}$&$124.7^{+27.0}_{-22.0}$&$134.2^{+10.7}_{-21.0}$&$-81.7^{+6.4}_{-5.7}$&$157.1^{+12.0}_{-21.1}$&$200.8^{+3.0}_{-3.1}$\\
NGC 6528&$0.7^{+0.1}_{-0.1}$&$139.5^{+11.6}_{-9.4}$&$340.5^{+5.6}_{-11.8}$&$-100.9^{+49.8}_{-33.7}$&$174.9^{+6.7}_{-19.2}$&$-98.7^{+15.8}_{-32.9}$&$204.3^{+15.7}_{-19.8}$&$228.3^{+0.5}_{-0.5}$\\
NGC 6535&$4.0^{+0.0}_{-0.0}$&$72.9^{+0.4}_{-0.4}$&$311.9^{+1.6}_{-1.7}$&$102.9^{+4.0}_{-4.1}$&$-8.5^{+1.9}_{-1.9}$&$84.1^{+3.6}_{-3.5}$&$84.5^{+3.8}_{-3.7}$&$133.2^{+1.0}_{-0.9}$\\
NGC 6539&$3.1^{+0.1}_{-0.1}$&$71.3^{+0.5}_{-0.2}$&$279.8^{+7.5}_{-7.1}$&$50.8^{+5.5}_{-6.9}$&$-172.3^{+6.2}_{-6.6}$&$-113.0^{+1.7}_{-2.8}$&$206.0^{+7.2}_{-6.1}$&$212.1^{+8.4}_{-7.5}$\\
NGC 6540&$2.3^{+0.2}_{-0.3}$&$97.9^{+1.4}_{-1.1}$&$351.5^{+1.2}_{-1.5}$&$6.1^{+0.6}_{-0.6}$&$-61.8^{+2.6}_{-2.8}$&$-120.2^{+5.6}_{-5.5}$&$135.2^{+3.8}_{-3.6}$&$135.4^{+3.8}_{-3.6}$\\
NGC 6541&$2.2^{+0.0}_{-0.0}$&$132.3^{+0.9}_{-0.9}$&$60.6^{+3.2}_{-3.0}$&$158.6^{+4.2}_{-4.4}$&$7.5^{+4.9}_{-3.6}$&$-200.1^{+5.2}_{-5.6}$&$200.3^{+5.8}_{-5.3}$&$255.5^{+1.9}_{-1.8}$\\
NGC 6544&$5.6^{+0.1}_{-0.1}$&$90.8^{+0.0}_{-0.0}$&$357.3^{+0.1}_{-0.1}$&$11.9^{+0.7}_{-0.7}$&$77.7^{+2.2}_{-2.0}$&$-22.6^{+5.1}_{-4.6}$&$80.9^{+0.8}_{-0.6}$&$81.8^{+0.7}_{-0.5}$\\
NGC 6553&$2.9^{+0.1}_{-0.1}$&$95.1^{+0.3}_{-0.3}$&$350.2^{+0.5}_{-0.5}$&$31.0^{+2.2}_{-2.1}$&$3.3^{+0.5}_{-0.5}$&$-232.2^{+0.9}_{-0.9}$&$232.2^{+0.9}_{-0.9}$&$234.3^{+0.7}_{-0.7}$\\
NGC 6558&$1.0^{+0.2}_{-0.1}$&$138.4^{+12.2}_{-10.0}$&$357.8^{+0.6}_{-1.3}$&$116.2^{+23.1}_{-32.3}$&$-148.5^{+22.6}_{-21.9}$&$-73.7^{+9.3}_{-7.7}$&$165.9^{+16.4}_{-16.2}$&$202.5^{+2.2}_{-2.1}$\\
NGC 6569&$2.6^{+0.3}_{-0.2}$&$117.3^{+2.2}_{-1.9}$&$182.2^{+0.2}_{-0.2}$&$-47.5^{+0.5}_{-0.5}$&$-1.4^{+2.3}_{-2.2}$&$-179.2^{+10.7}_{-10.8}$&$179.2^{+10.9}_{-10.7}$&$185.4^{+10.5}_{-10.2}$\\
NGC 6584&$7.0^{+0.1}_{-0.1}$&$123.0^{+0.3}_{-0.3}$&$137.0^{+0.6}_{-0.7}$&$302.7^{+1.5}_{-1.5}$&$87.0^{+3.3}_{-4.0}$&$-121.1^{+7.9}_{-7.4}$&$149.1^{+7.6}_{-8.6}$&$337.5^{+4.3}_{-4.7}$\\
NGC 6624&$1.2^{+0.0}_{-0.0}$&$158.4^{+1.5}_{-2.4}$&$295.0^{+11.3}_{-13.8}$&$107.9^{+7.1}_{-7.6}$&$60.8^{+11.5}_{-12.3}$&$-56.7^{+2.5}_{-1.6}$&$83.2^{+7.1}_{-7.2}$&$136.4^{+1.4}_{-1.3}$\\
NGC 6626&$3.0^{+0.1}_{-0.1}$&$99.7^{+0.4}_{-0.4}$&$345.6^{+0.6}_{-0.6}$&$-13.0^{+0.5}_{-0.6}$&$95.2^{+2.0}_{-1.7}$&$-45.8^{+2.6}_{-2.3}$&$105.7^{+0.9}_{-0.8}$&$106.5^{+0.8}_{-0.7}$\\
NGC 6637&$1.7^{+0.1}_{-0.0}$&$156.6^{+2.3}_{-2.6}$&$203.0^{+3.1}_{-2.7}$&$-56.4^{+4.3}_{-3.9}$&$-75.5^{+6.9}_{-6.2}$&$-97.6^{+1.1}_{-1.4}$&$123.2^{+4.9}_{-4.6}$&$135.4^{+2.9}_{-2.5}$\\
NGC 6638&$2.3^{+0.2}_{-0.3}$&$121.2^{+3.2}_{-2.1}$&$222.6^{+7.3}_{-4.2}$&$32.7^{+1.3}_{-1.2}$&$-49.4^{+2.6}_{-3.3}$&$-18.1^{+1.4}_{-1.7}$&$52.4^{+3.3}_{-2.1}$&$61.9^{+2.9}_{-2.2}$\\
NGC 6642&$1.6^{+0.0}_{-0.0}$&$122.6^{+0.2}_{-0.5}$&$276.9^{+6.4}_{-6.0}$&$107.2^{+3.5}_{-4.1}$&$-12.1^{+3.4}_{-2.3}$&$21.4^{+9.9}_{-11.0}$&$24.8^{+7.9}_{-6.6}$&$110.1^{+2.1}_{-2.0}$\\
NGC 6652&$2.2^{+0.1}_{-0.1}$&$147.4^{+2.8}_{-2.5}$&$192.2^{+1.4}_{-1.1}$&$-168.3^{+0.9}_{-0.7}$&$-41.8^{+9.3}_{-8.7}$&$-28.5^{+5.7}_{-5.1}$&$50.5^{+10.4}_{-10.8}$&$175.9^{+2.9}_{-2.9}$\\
NGC 6656&$4.9^{+0.0}_{-0.0}$&$94.7^{+0.1}_{-0.1}$&$353.5^{+0.1}_{-0.1}$&$189.9^{+1.1}_{-1.0}$&$133.8^{+1.7}_{-1.7}$&$-181.5^{+0.6}_{-0.5}$&$225.5^{+0.7}_{-0.7}$&$294.8^{+1.1}_{-1.1}$\\
NGC 6681&$2.3^{+0.1}_{-0.1}$&$151.2^{+2.1}_{-1.9}$&$204.4^{+2.5}_{-1.9}$&$262.2^{+4.0}_{-5.1}$&$-101.2^{+10.1}_{-11.2}$&$-5.9^{+5.6}_{-7.4}$&$101.4^{+11.8}_{-10.3}$&$281.1^{+0.7}_{-0.6}$\\
NGC 6712&$3.5^{+0.0}_{-0.0}$&$98.7^{+0.2}_{-0.3}$&$294.0^{+4.1}_{-3.7}$&$149.2^{+1.9}_{-2.3}$&$139.8^{+5.1}_{-5.7}$&$11.4^{+8.2}_{-9.7}$&$140.3^{+6.0}_{-6.1}$&$204.8^{+2.7}_{-2.9}$\\
NGC 6715&$18.6^{+0.3}_{-0.3}$&$110.1^{+0.1}_{-0.1}$&$188.2^{+0.0}_{-0.0}$&$144.0^{+0.4}_{-0.4}$&$-268.5^{+4.1}_{-4.0}$&$-63.8^{+3.8}_{-4.1}$&$276.0^{+4.5}_{-4.7}$&$311.2^{+3.9}_{-4.0}$\\
NGC 6717&$2.3^{+0.0}_{-0.0}$&$126.5^{+0.7}_{-0.7}$&$299.3^{+2.6}_{-2.7}$&$-20.2^{+1.5}_{-1.5}$&$-17.7^{+1.4}_{-1.4}$&$-100.2^{+1.1}_{-1.1}$&$101.7^{+1.1}_{-0.9}$&$103.7^{+1.2}_{-1.1}$\\
NGC 6723&$2.4^{+0.0}_{-0.0}$&$174.7^{+1.7}_{-1.6}$&$357.6^{+0.6}_{-1.1}$&$43.6^{+2.2}_{-2.4}$&$-104.8^{+3.0}_{-4.6}$&$-161.6^{+2.6}_{-2.1}$&$192.7^{+0.9}_{-0.9}$&$197.7^{+0.8}_{-1.0}$\\
NGC 6749&$4.9^{+0.1}_{-0.0}$&$93.1^{+0.1}_{-0.1}$&$293.9^{+2.5}_{-2.4}$&$-30.9^{+2.4}_{-2.5}$&$0.8^{+0.8}_{-0.8}$&$-100.1^{+3.0}_{-3.0}$&$100.1^{+3.0}_{-3.0}$&$104.8^{+3.5}_{-3.5}$\\
NGC 6752&$5.2^{+0.0}_{-0.0}$&$109.6^{+0.3}_{-0.3}$&$17.4^{+0.3}_{-0.3}$&$-39.9^{+0.7}_{-0.7}$&$-49.7^{+0.2}_{-0.3}$&$-161.4^{+0.9}_{-0.9}$&$168.8^{+0.9}_{-0.8}$&$173.5^{+0.7}_{-0.7}$\\
NGC 6760&$5.1^{+0.2}_{-0.1}$&$96.1^{+0.1}_{-0.2}$&$285.9^{+4.6}_{-4.1}$&$90.1^{+2.3}_{-3.2}$&$9.8^{+1.2}_{-1.1}$&$-122.3^{+3.9}_{-5.0}$&$122.7^{+5.0}_{-3.8}$&$152.2^{+2.5}_{-2.0}$\\
NGC 6779&$9.9^{+0.1}_{-0.1}$&$81.1^{+0.0}_{-0.0}$&$290.2^{+0.6}_{-0.6}$&$153.4^{+0.7}_{-0.7}$&$-88.2^{+2.1}_{-2.1}$&$39.2^{+1.8}_{-2.0}$&$96.5^{+2.6}_{-2.6}$&$181.3^{+1.5}_{-1.5}$\\
NGC 6809&$3.9^{+0.0}_{-0.0}$&$122.0^{+0.6}_{-0.5}$&$347.0^{+0.3}_{-0.3}$&$-140.6^{+1.2}_{-1.1}$&$153.7^{+1.7}_{-1.6}$&$-62.8^{+1.2}_{-1.1}$&$166.0^{+1.2}_{-1.1}$&$217.6^{+0.2}_{-0.2}$\\
NGC 6838&$6.8^{+0.0}_{-0.0}$&$92.5^{+0.0}_{-0.0}$&$330.7^{+0.4}_{-0.3}$&$32.3^{+0.5}_{-0.4}$&$-40.2^{+0.4}_{-0.4}$&$-188.7^{+0.3}_{-0.4}$&$192.9^{+0.3}_{-0.3}$&$195.6^{+0.3}_{-0.3}$\\
NGC 6864&$14.3^{+0.4}_{-0.4}$&$128.4^{+0.3}_{-0.3}$&$214.8^{+0.5}_{-0.5}$&$-115.5^{+1.3}_{-1.2}$&$25.7^{+2.2}_{-2.1}$&$-29.2^{+6.9}_{-7.0}$&$39.0^{+6.1}_{-5.8}$&$121.9^{+3.1}_{-2.6}$\\
NGC 6934&$12.8^{+0.1}_{-0.1}$&$113.3^{+0.0}_{-0.0}$&$265.1^{+0.4}_{-0.4}$&$-328.9^{+1.8}_{-1.7}$&$8.7^{+1.5}_{-1.5}$&$-113.5^{+6.0}_{-6.3}$&$113.8^{+6.3}_{-6.1}$&$348.0^{+3.7}_{-3.6}$\\
NGC 6981&$12.5^{+0.2}_{-0.1}$&$135.7^{+0.1}_{-0.1}$&$247.5^{+0.4}_{-0.5}$&$-241.1^{+1.5}_{-1.5}$&$2.1^{+2.0}_{-1.8}$&$-2.4^{+2.9}_{-5.1}$&$4.1^{+4.3}_{-2.7}$&$241.2^{+1.5}_{-1.6}$\\
NGC 7006&$36.7^{+0.4}_{-0.4}$&$110.8^{+0.0}_{-0.0}$&$256.0^{+0.2}_{-0.1}$&$-178.1^{+1.1}_{-1.0}$&$-22.7^{+4.8}_{-4.4}$&$22.7^{+4.6}_{-4.9}$&$32.4^{+5.4}_{-5.6}$&$181.1^{+0.9}_{-0.8}$\\
NGC 7078&$10.7^{+0.1}_{-0.1}$&$117.1^{+0.1}_{-0.1}$&$295.4^{+0.4}_{-0.3}$&$7.0^{+0.8}_{-0.8}$&$30.3^{+1.3}_{-1.3}$&$-113.1^{+1.2}_{-1.1}$&$117.1^{+1.3}_{-1.4}$&$117.3^{+1.3}_{-1.3}$\\
NGC 7089&$10.5^{+0.1}_{-0.1}$&$130.4^{+0.1}_{-0.1}$&$288.0^{+0.5}_{-0.5}$&$227.6^{+0.9}_{-0.8}$&$33.5^{+2.0}_{-1.9}$&$28.9^{+2.6}_{-2.4}$&$44.3^{+3.1}_{-2.9}$&$231.8^{+1.4}_{-1.2}$\\
NGC 7099&$7.3^{+0.0}_{-0.0}$&$147.1^{+0.3}_{-0.3}$&$318.3^{+0.6}_{-0.6}$&$-122.8^{+1.4}_{-1.4}$&$-8.7^{+3.0}_{-3.0}$&$70.9^{+0.9}_{-0.9}$&$71.5^{+0.8}_{-0.7}$&$142.1^{+1.2}_{-1.2}$\\
NGC 7492&$23.6^{+0.5}_{-0.5}$&$157.8^{+0.1}_{-0.1}$&$280.5^{+1.2}_{-1.1}$&$-98.4^{+0.8}_{-0.8}$&$52.5^{+7.7}_{-7.5}$&$13.6^{+2.4}_{-2.5}$&$54.3^{+7.1}_{-6.9}$&$112.4^{+3.7}_{-3.5}$\\
Pal 1&$17.4^{+0.2}_{-0.2}$&$77.8^{+0.1}_{-0.1}$&$331.5^{+0.3}_{-0.3}$&$31.2^{+1.0}_{-1.0}$&$32.4^{+1.5}_{-1.4}$&$-200.6^{+1.8}_{-1.6}$&$203.2^{+1.7}_{-1.8}$&$205.6^{+1.6}_{-1.7}$\\
Pal 10&$7.6^{+0.7}_{-0.6}$&$86.6^{+0.2}_{-0.1}$&$289.9^{+8.0}_{-6.8}$&$-96.5^{+7.1}_{-4.6}$&$-32.0^{+3.4}_{-3.3}$&$-256.6^{+34.7}_{-39.8}$&$258.6^{+39.8}_{-34.9}$&$275.9^{+39.1}_{-35.0}$\\
Pal 11&$8.7^{+0.4}_{-0.4}$&$115.4^{+0.3}_{-0.3}$&$244.8^{+2.0}_{-1.8}$&$-23.1^{+1.5}_{-1.4}$&$25.3^{+2.3}_{-2.3}$&$-157.7^{+10.9}_{-11.3}$&$159.9^{+11.3}_{-11.1}$&$161.6^{+11.3}_{-11.3}$\\
Pal 12&$15.3^{+0.3}_{-0.3}$&$153.4^{+0.2}_{-0.2}$&$247.6^{+1.0}_{-1.0}$&$-39.6^{+0.9}_{-0.9}$&$-163.8^{+4.7}_{-4.8}$&$-300.0^{+4.3}_{-4.7}$&$341.7^{+6.1}_{-5.9}$&$344.0^{+6.1}_{-5.9}$\\
Pal 13&$24.5^{+0.3}_{-0.3}$&$130.4^{+0.1}_{-0.1}$&$292.8^{+0.3}_{-0.3}$&$242.2^{+1.8}_{-1.5}$&$-101.8^{+4.8}_{-4.8}$&$71.0^{+5.8}_{-6.0}$&$124.0^{+5.8}_{-5.7}$&$272.2^{+4.1}_{-3.9}$\\
Pal 14&$68.7^{+1.1}_{-1.2}$&$43.9^{+0.1}_{-0.1}$&$213.5^{+0.1}_{-0.1}$&$169.8^{+0.7}_{-0.6}$&$-0.3^{+7.6}_{-9.6}$&$-0.0^{+9.1}_{-9.2}$&$9.6^{+10.2}_{-6.4}$&$170.0^{+1.4}_{-0.5}$\\
Pal 15&$37.4^{+0.9}_{-0.8}$&$60.8^{+0.1}_{-0.1}$&$203.5^{+0.1}_{-0.1}$&$151.8^{+1.6}_{-1.4}$&$30.2^{+6.8}_{-7.2}$&$-13.1^{+6.1}_{-7.2}$&$33.9^{+7.0}_{-7.7}$&$155.7^{+1.5}_{-1.5}$\\
Pal 2&$34.1^{+0.8}_{-1.3}$&$96.9^{+0.1}_{-0.1}$&$352.8^{+0.1}_{-0.1}$&$-108.4^{+1.0}_{-1.2}$&$4.3^{+3.4}_{-3.1}$&$-6.1^{+7.1}_{-12.7}$&$10.1^{+9.9}_{-6.1}$&$109.1^{+1.9}_{-1.4}$\\
Pal 3&$98.1^{+2.4}_{-2.1}$&$49.8^{+0.0}_{-0.0}$&$54.7^{+0.1}_{-0.1}$&$-61.0^{+2.0}_{-1.9}$&$-130.8^{+20.6}_{-20.3}$&$-48.8^{+27.0}_{-26.3}$&$141.7^{+22.4}_{-22.9}$&$154.5^{+20.3}_{-21.1}$\\
Pal 4&$104.1^{+1.7}_{-1.7}$&$22.2^{+0.1}_{-0.1}$&$17.8^{+0.1}_{-0.1}$&$46.6^{+1.3}_{-1.6}$&$-12.7^{+13.6}_{-21.1}$&$11.1^{+18.9}_{-12.9}$&$23.3^{+19.6}_{-16.3}$&$52.2^{+10.1}_{-4.0}$\\
Pal 5&$17.3^{+0.5}_{-0.5}$&$24.4^{+0.6}_{-0.6}$&$181.8^{+0.1}_{-0.0}$&$-33.2^{+0.8}_{-0.7}$&$-43.5^{+1.9}_{-2.0}$&$-153.9^{+9.7}_{-10.0}$&$159.8^{+9.7}_{-9.4}$&$163.3^{+9.5}_{-9.3}$\\
Pal 6&$1.0^{+0.4}_{-0.3}$&$75.9^{+4.7}_{-8.9}$&$344.6^{+5.4}_{-11.4}$&$-154.3^{+33.5}_{-13.6}$&$-229.0^{+25.2}_{-33.0}$&$16.1^{+4.4}_{-22.7}$&$230.0^{+32.6}_{-25.3}$&$277.0^{+11.7}_{-11.9}$\\
Pal 7&$4.3^{+0.1}_{-0.1}$&$83.6^{+0.5}_{-0.4}$&$336.7^{+1.8}_{-1.7}$&$-99.7^{+4.9}_{-5.0}$&$-35.3^{+0.5}_{-0.5}$&$-254.0^{+0.6}_{-0.6}$&$256.4^{+0.6}_{-0.6}$&$275.0^{+2.0}_{-1.8}$\\
Pal 8&$4.1^{+0.5}_{-0.5}$&$108.6^{+1.5}_{-1.3}$&$224.4^{+5.2}_{-4.1}$&$-8.0^{+2.0}_{-1.5}$&$33.8^{+2.7}_{-2.4}$&$-95.1^{+16.4}_{-16.8}$&$100.9^{+16.8}_{-16.0}$&$101.3^{+16.8}_{-16.1}$\\
Pyxis&$38.6^{+0.5}_{-0.5}$&$83.3^{+0.0}_{-0.0}$&$69.2^{+0.2}_{-0.2}$&$-207.7^{+0.9}_{-0.9}$&$-184.8^{+6.6}_{-6.4}$&$28.5^{+4.5}_{-4.3}$&$187.1^{+6.4}_{-6.9}$&$279.5^{+4.6}_{-4.8}$\\
Rup 106&$18.1^{+0.3}_{-0.2}$&$76.5^{+0.0}_{-0.0}$&$97.5^{+0.3}_{-0.3}$&$-214.0^{+0.9}_{-0.9}$&$-83.9^{+2.1}_{-2.3}$&$-90.9^{+2.0}_{-1.9}$&$123.8^{+2.3}_{-2.4}$&$247.2^{+0.7}_{-0.7}$\\
Segue 3&$27.9^{+0.9}_{-0.8}$&$112.5^{+0.0}_{-0.0}$&$266.5^{+0.5}_{-0.5}$&$-12.2^{+3.6}_{-3.3}$&$-19.4^{+11.9}_{-12.1}$&$-198.7^{+14.8}_{-16.0}$&$200.0^{+16.7}_{-15.0}$&$200.3^{+16.9}_{-15.0}$\\
Terzan 1&$2.5^{+0.2}_{-0.2}$&$87.1^{+0.2}_{-0.3}$&$5.6^{+0.6}_{-0.5}$&$-68.5^{+1.0}_{-1.1}$&$-4.3^{+1.0}_{-1.1}$&$-80.7^{+5.5}_{-5.4}$&$80.8^{+5.4}_{-5.4}$&$106.0^{+3.8}_{-4.1}$\\
Terzan 10&$2.3^{+0.4}_{-0.4}$&$97.7^{+1.0}_{-0.8}$&$200.2^{+3.0}_{-2.2}$&$195.2^{+4.6}_{-6.1}$&$-257.4^{+6.6}_{-6.1}$&$-103.0^{+1.9}_{-2.6}$&$277.1^{+6.0}_{-5.3}$&$338.9^{+7.5}_{-8.0}$\\
Terzan 12&$3.1^{+0.3}_{-0.2}$&$93.0^{+0.5}_{-0.5}$&$346.1^{+2.2}_{-2.0}$&$-101.4^{+1.5}_{-1.2}$&$-95.7^{+5.7}_{-4.4}$&$-144.8^{+3.5}_{-5.4}$&$174.0^{+1.4}_{-1.0}$&$201.3^{+1.7}_{-1.2}$\\
Terzan 2&$0.8^{+0.2}_{-0.1}$&$63.9^{+5.9}_{-5.3}$&$46.5^{+18.5}_{-13.7}$&$-101.4^{+34.7}_{-22.2}$&$6.0^{+12.3}_{-2.9}$&$96.7^{+25.8}_{-30.5}$&$96.8^{+27.5}_{-30.4}$&$140.4^{+0.9}_{-0.8}$\\
Terzan 3&$2.5^{+0.0}_{-0.0}$&$59.8^{+1.6}_{-1.3}$&$65.2^{+8.0}_{-7.0}$&$30.2^{+12.5}_{-14.7}$&$-84.0^{+11.6}_{-14.0}$&$-208.5^{+3.8}_{-1.7}$&$224.8^{+2.3}_{-2.5}$&$226.9^{+0.8}_{-0.7}$\\
Terzan 4&$0.7^{+0.2}_{-0.1}$&$73.0^{+4.4}_{-3.0}$&$54.2^{+23.0}_{-18.2}$&$44.8^{+3.1}_{-9.7}$&$-89.9^{+3.7}_{-5.7}$&$-57.4^{+1.7}_{-1.5}$&$106.7^{+5.6}_{-3.8}$&$115.8^{+2.3}_{-2.4}$\\
Terzan 5&$1.6^{+0.1}_{-0.1}$&$82.1^{+0.7}_{-0.9}$&$343.8^{+1.5}_{-1.9}$&$70.3^{+0.9}_{-1.0}$&$36.9^{+1.9}_{-2.0}$&$-42.9^{+6.7}_{-6.0}$&$56.6^{+3.8}_{-4.0}$&$90.1^{+3.1}_{-3.0}$\\
Terzan 6&$0.8^{+0.3}_{-0.3}$&$107.2^{+10.1}_{-5.3}$&$13.0^{+8.9}_{-4.2}$&$-119.2^{+24.4}_{-10.5}$&$16.8^{+8.6}_{-8.4}$&$103.6^{+29.7}_{-22.6}$&$105.1^{+31.0}_{-23.6}$&$158.8^{+6.5}_{-5.9}$\\
Terzan 7&$16.9^{+0.5}_{-0.5}$&$119.4^{+0.3}_{-0.3}$&$185.3^{+0.1}_{-0.1}$&$131.0^{+0.5}_{-0.5}$&$-300.8^{+5.5}_{-5.8}$&$-65.6^{+6.6}_{-6.5}$&$307.8^{+7.1}_{-6.5}$&$334.5^{+6.4}_{-6.0}$\\
Terzan 8&$20.5^{+0.4}_{-0.4}$&$123.9^{+0.2}_{-0.2}$&$188.5^{+0.1}_{-0.1}$&$127.6^{+0.5}_{-0.5}$&$-293.7^{+4.7}_{-5.1}$&$-68.7^{+4.7}_{-5.1}$&$301.8^{+5.7}_{-5.5}$&$327.7^{+5.2}_{-4.9}$\\
Terzan 9&$2.5^{+0.3}_{-0.3}$&$93.9^{+0.9}_{-0.7}$&$351.9^{+1.4}_{-1.8}$&$-80.1^{+0.5}_{-0.5}$&$54.2^{+4.2}_{-4.3}$&$-47.9^{+9.8}_{-10.7}$&$72.7^{+4.4}_{-2.8}$&$108.1^{+2.9}_{-1.6}$\\
Ton 2&$1.7^{+0.3}_{-0.3}$&$103.4^{+3.8}_{-2.5}$&$42.7^{+18.5}_{-9.4}$&$-2.9^{+41.8}_{-78.0}$&$-172.2^{+7.2}_{-1.9}$&$-227.5^{+4.4}_{-2.5}$&$285.4^{+2.8}_{-7.3}$&$288.2^{+2.8}_{-2.0}$\\
UKS 1&$7.6^{+0.5}_{-0.4}$&$88.2^{+0.1}_{-0.1}$&$190.6^{+0.3}_{-0.3}$&$89.7^{+2.0}_{-2.0}$&$-32.5^{+6.0}_{-5.9}$&$-22.5^{+8.1}_{-9.7}$&$40.2^{+8.6}_{-8.6}$&$98.4^{+3.8}_{-3.2}$\\
VVV CL001&$1.1^{+0.5}_{-0.2}$&$82.9^{+2.1}_{-2.5}$&$302.3^{+22.7}_{-82.7}$&$252.8^{+63.3}_{-429.0}$&$-48.6^{+7.1}_{-68.8}$&$185.5^{+97.5}_{-91.6}$&$202.2^{+94.9}_{-98.2}$&$330.3^{+2.6}_{-2.9}$\\
Whiting 1&$35.1^{+1.0}_{-1.0}$&$139.4^{+0.3}_{-0.3}$&$348.1^{+0.1}_{-0.1}$&$-147.7^{+2.1}_{-2.0}$&$185.5^{+15.0}_{-12.4}$&$-66.9^{+7.2}_{-7.2}$&$197.4^{+14.0}_{-11.1}$&$246.6^{+12.3}_{-10.0}$\\

\end{longtable}

\begin{longtable}[c]{lccc|lccc}
\label{tab:dyn}\\
\caption{Orbital properties of globular clusters for Model PNFW.}\\
\hline
\hline
name & $R_\mathrm{peri}$ & $R_\mathrm{apo}$ & $ecc$ & name & $R_\mathrm{peri}$ & $R_\mathrm{apo}$ & $ecc$\\
	 & (kpc) & (kpc) & & & (kpc) & (kpc) & \\
\hline
\vspace{-10pt}
\endhead
\hline
\endfoot
\hline
\endlastfoot
NGC 104&$5.5^{+0.0}_{-0.0}$&$7.5^{+0.0}_{-0.0}$&$0.15^{+0.0}_{-0.0}$&NGC 6366&$1.9^{+0.0}_{-0.0}$&$5.7^{+0.1}_{-0.0}$&$0.5^{+0.0}_{-0.0}$\\
NGC 288&$2.8^{+0.1}_{-0.1}$&$12.3^{+0.1}_{-0.1}$&$0.63^{+0.01}_{-0.01}$&Terzan 4&$0.2^{+0.0}_{-0.0}$&$0.7^{+0.2}_{-0.1}$&$0.57^{+0.03}_{-0.03}$\\
NGC 362&$0.4^{+0.1}_{-0.1}$&$12.6^{+0.2}_{-0.2}$&$0.93^{+0.01}_{-0.01}$&BH 229&$0.9^{+0.1}_{-0.1}$&$1.7^{+0.1}_{-0.1}$&$0.29^{+0.02}_{-0.01}$\\
Whiting 1&$22.5^{+2.1}_{-1.7}$&$80.1^{+14.5}_{-8.9}$&$0.56^{+0.03}_{-0.01}$&FSR 1758&$3.2^{+0.8}_{-0.8}$&$12.0^{+5.1}_{-3.5}$&$0.58^{+0.04}_{-0.02}$\\
NGC 1261&$1.3^{+0.2}_{-0.1}$&$21.1^{+0.2}_{-0.2}$&$0.88^{+0.01}_{-0.01}$&NGC 6362&$3.0^{+0.0}_{-0.0}$&$5.2^{+0.0}_{-0.0}$&$0.27^{+0.0}_{-0.0}$\\
Pal 1&$14.9^{+0.3}_{-0.3}$&$18.9^{+0.4}_{-0.3}$&$0.12^{+0.01}_{-0.0}$&Liller 1&$0.1^{+0.0}_{-0.0}$&$0.8^{+0.1}_{-0.0}$&$0.79^{+0.09}_{-0.09}$\\
E 1&$12.3^{+23.9}_{-9.3}$&$126.7^{+5.6}_{-3.9}$&$0.82^{+0.13}_{-0.26}$&NGC 6380&$0.2^{+0.0}_{-0.0}$&$2.2^{+0.2}_{-0.2}$&$0.86^{+0.03}_{-0.01}$\\
Eridanus&$15.9^{+5.4}_{-5.0}$&$144.5^{+8.6}_{-5.9}$&$0.8^{+0.05}_{-0.05}$&Terzan 1&$0.5^{+0.1}_{-0.1}$&$2.6^{+0.2}_{-0.2}$&$0.69^{+0.03}_{-0.03}$\\
Pal 2&$0.6^{+0.6}_{-0.1}$&$39.9^{+1.0}_{-1.5}$&$0.97^{+0.01}_{-0.03}$&Ton 2&$1.7^{+0.2}_{-0.4}$&$2.9^{+0.5}_{-0.3}$&$0.27^{+0.05}_{-0.01}$\\
NGC 1851&$1.6^{+0.0}_{-0.0}$&$21.0^{+0.1}_{-0.1}$&$0.86^{+0.0}_{-0.0}$&NGC 6388&$1.0^{+0.0}_{-0.0}$&$4.0^{+0.1}_{-0.1}$&$0.59^{+0.01}_{-0.01}$\\
NGC 1904&$0.2^{+0.0}_{-0.0}$&$19.6^{+0.2}_{-0.1}$&$0.98^{+0.0}_{-0.0}$&NGC 6402&$0.5^{+0.1}_{-0.1}$&$4.0^{+0.1}_{-0.1}$&$0.77^{+0.02}_{-0.02}$\\
NGC 2298&$1.6^{+0.1}_{-0.1}$&$16.4^{+0.2}_{-0.2}$&$0.82^{+0.01}_{-0.01}$&NGC 6401&$0.2^{+0.1}_{-0.1}$&$1.5^{+0.1}_{-0.0}$&$0.73^{+0.07}_{-0.1}$\\
NGC 2419&$13.4^{+4.0}_{-4.1}$&$97.5^{+1.7}_{-1.9}$&$0.76^{+0.07}_{-0.06}$&NGC 6397&$2.5^{+0.0}_{-0.0}$&$6.3^{+0.0}_{-0.0}$&$0.43^{+0.0}_{-0.0}$\\
Pyxis&$19.9^{+1.0}_{-1.0}$&$142.8^{+12.3}_{-10.9}$&$0.75^{+0.01}_{-0.01}$&Pal 6&$0.6^{+0.1}_{-0.1}$&$1.9^{+0.7}_{-0.8}$&$0.51^{+0.05}_{-0.11}$\\
NGC 2808&$0.8^{+0.0}_{-0.0}$&$14.6^{+0.1}_{-0.1}$&$0.9^{+0.0}_{-0.0}$&NGC 6426&$4.1^{+0.5}_{-0.4}$&$18.1^{+0.7}_{-0.6}$&$0.63^{+0.02}_{-0.02}$\\
E 3&$8.9^{+0.2}_{-0.1}$&$13.3^{+1.0}_{-0.9}$&$0.2^{+0.03}_{-0.03}$&Djorg 1&$0.9^{+0.6}_{-0.2}$&$6.8^{+7.1}_{-2.4}$&$0.77^{+0.04}_{-0.05}$\\
Pal 3&$63.2^{+14.4}_{-16.4}$&$123.7^{+22.1}_{-9.7}$&$0.34^{+0.09}_{-0.03}$&Terzan 5&$0.1^{+0.0}_{-0.0}$&$1.7^{+0.1}_{-0.1}$&$0.85^{+0.02}_{-0.02}$\\
NGC 3201&$8.4^{+0.0}_{-0.0}$&$33.8^{+0.6}_{-0.5}$&$0.6^{+0.0}_{-0.0}$&NGC 6440&$0.1^{+0.0}_{-0.0}$&$1.3^{+0.0}_{-0.0}$&$0.8^{+0.02}_{-0.02}$\\
ESO 93-8&$12.4^{+0.9}_{-1.7}$&$13.1^{+2.0}_{-0.6}$&$0.05^{+0.04}_{-0.03}$&NGC 6441&$1.8^{+0.1}_{-0.1}$&$4.9^{+0.1}_{-0.2}$&$0.46^{+0.02}_{-0.02}$\\
Pal 4&$5.3^{+6.0}_{-3.9}$&$108.9^{+1.9}_{-1.9}$&$0.91^{+0.07}_{-0.09}$&Terzan 6&$0.2^{+0.0}_{-0.0}$&$1.0^{+0.4}_{-0.4}$&$0.66^{+0.09}_{-0.15}$\\
Crater&$73.8^{+71.9}_{-58.4}$&$148.7^{+55.0}_{-3.7}$&$0.5^{+0.4}_{-0.34}$&NGC 6453&$0.8^{+0.1}_{-0.1}$&$2.4^{+0.2}_{-0.2}$&$0.5^{+0.01}_{-0.01}$\\
NGC 4147&$1.3^{+0.1}_{-0.1}$&$26.4^{+0.2}_{-0.3}$&$0.91^{+0.01}_{-0.01}$&UKS 1&$0.3^{+0.2}_{-0.1}$&$8.2^{+0.5}_{-0.5}$&$0.93^{+0.03}_{-0.04}$\\
NGC 4372&$2.8^{+0.1}_{-0.1}$&$7.2^{+0.1}_{-0.1}$&$0.45^{+0.01}_{-0.01}$&VVV CL001&$0.4^{+0.6}_{-0.2}$&$3.3^{+1.5}_{-1.0}$&$0.73^{+0.15}_{-0.19}$\\
Rup 106&$5.0^{+0.2}_{-0.2}$&$36.9^{+0.6}_{-0.6}$&$0.76^{+0.01}_{-0.01}$&NGC 6496&$2.7^{+0.1}_{-0.1}$&$5.3^{+0.3}_{-0.3}$&$0.32^{+0.01}_{-0.01}$\\
NGC 4590&$9.1^{+0.1}_{-0.0}$&$33.6^{+0.5}_{-0.5}$&$0.57^{+0.0}_{-0.0}$&Terzan 9&$0.3^{+0.1}_{-0.1}$&$2.7^{+0.4}_{-0.4}$&$0.83^{+0.02}_{-0.04}$\\
BH 140&$1.9^{+0.1}_{-0.1}$&$10.6^{+0.9}_{-0.7}$&$0.7^{+0.0}_{-0.0}$&Djorg 2&$0.6^{+0.1}_{-0.3}$&$0.8^{+0.2}_{-0.1}$&$0.16^{+0.15}_{-0.04}$\\
NGC 4833&$0.6^{+0.0}_{-0.0}$&$8.1^{+0.1}_{-0.1}$&$0.86^{+0.01}_{-0.01}$&NGC 6517&$0.3^{+0.0}_{-0.0}$&$3.3^{+0.2}_{-0.2}$&$0.82^{+0.01}_{-0.02}$\\
NGC 5024&$8.7^{+0.2}_{-0.2}$&$23.2^{+0.2}_{-0.2}$&$0.45^{+0.01}_{-0.01}$&Terzan 10&$1.6^{+0.3}_{-0.3}$&$6.3^{+1.3}_{-1.1}$&$0.59^{+0.01}_{-0.01}$\\
NGC 5053&$10.8^{+0.2}_{-0.2}$&$18.1^{+0.2}_{-0.2}$&$0.25^{+0.01}_{-0.01}$&NGC 6522&$0.4^{+0.2}_{-0.1}$&$1.1^{+0.2}_{-0.1}$&$0.48^{+0.1}_{-0.08}$\\
NGC 5139&$2.0^{+0.0}_{-0.0}$&$6.9^{+0.0}_{-0.0}$&$0.55^{+0.01}_{-0.01}$&NGC 6535&$1.0^{+0.1}_{-0.1}$&$4.5^{+0.1}_{-0.1}$&$0.63^{+0.02}_{-0.02}$\\
NGC 5272&$4.9^{+0.1}_{-0.0}$&$16.4^{+0.1}_{-0.1}$&$0.54^{+0.0}_{-0.0}$&NGC 6528&$0.4^{+0.0}_{-0.0}$&$0.9^{+0.3}_{-0.2}$&$0.38^{+0.13}_{-0.15}$\\
NGC 5286&$0.7^{+0.1}_{-0.1}$&$12.9^{+0.1}_{-0.1}$&$0.9^{+0.01}_{-0.01}$&NGC 6539&$2.5^{+0.2}_{-0.2}$&$3.4^{+0.2}_{-0.1}$&$0.15^{+0.02}_{-0.01}$\\
AM 4&$24.3^{+0.8}_{-3.2}$&$40.2^{+45.4}_{-15.3}$&$0.29^{+0.28}_{-0.2}$&NGC 6540&$0.9^{+0.2}_{-0.2}$&$2.3^{+0.2}_{-0.3}$&$0.44^{+0.04}_{-0.03}$\\
NGC 5466&$6.8^{+0.3}_{-0.3}$&$64.0^{+3.2}_{-2.9}$&$0.81^{+0.0}_{-0.0}$&NGC 6544&$0.2^{+0.1}_{-0.0}$&$5.6^{+0.1}_{-0.1}$&$0.91^{+0.02}_{-0.02}$\\
NGC 5634&$3.3^{+0.5}_{-0.4}$&$22.4^{+0.5}_{-0.5}$&$0.74^{+0.03}_{-0.03}$&NGC 6541&$1.3^{+0.1}_{-0.0}$&$3.8^{+0.0}_{-0.0}$&$0.49^{+0.01}_{-0.01}$\\
NGC 5694&$2.9^{+0.3}_{-0.3}$&$70.8^{+2.0}_{-2.0}$&$0.92^{+0.01}_{-0.01}$&ESO 280-06&$2.2^{+0.4}_{-0.3}$&$14.1^{+0.7}_{-0.6}$&$0.73^{+0.02}_{-0.03}$\\
IC 4499&$7.0^{+0.2}_{-0.2}$&$29.9^{+0.6}_{-0.6}$&$0.62^{+0.0}_{-0.0}$&NGC 6553&$2.7^{+0.1}_{-0.1}$&$3.5^{+0.1}_{-0.1}$&$0.12^{+0.01}_{-0.01}$\\
Munoz 1&$30.9^{+14.0}_{-12.8}$&$48.9^{+5.3}_{-2.1}$&$0.24^{+0.21}_{-0.14}$&NGC 6558&$0.4^{+0.0}_{-0.0}$&$1.3^{+0.4}_{-0.3}$&$0.51^{+0.1}_{-0.13}$\\
NGC 5824&$16.3^{+1.2}_{-1.2}$&$39.7^{+2.6}_{-2.4}$&$0.42^{+0.01}_{-0.0}$&Pal 7&$3.6^{+0.1}_{-0.1}$&$7.0^{+0.3}_{-0.3}$&$0.32^{+0.01}_{-0.01}$\\
Pal 5&$10.8^{+1.5}_{-1.3}$&$17.9^{+0.7}_{-0.6}$&$0.25^{+0.04}_{-0.04}$&Terzan 12&$1.6^{+0.2}_{-0.1}$&$3.8^{+0.4}_{-0.3}$&$0.39^{+0.0}_{-0.0}$\\
NGC 5897&$2.5^{+0.3}_{-0.3}$&$9.5^{+0.4}_{-0.4}$&$0.59^{+0.03}_{-0.03}$&NGC 6569&$1.9^{+0.3}_{-0.3}$&$2.8^{+0.3}_{-0.3}$&$0.19^{+0.03}_{-0.03}$\\
NGC 5904&$2.7^{+0.0}_{-0.0}$&$28.3^{+0.8}_{-0.8}$&$0.82^{+0.0}_{-0.0}$&BH 261&$1.6^{+0.2}_{-0.2}$&$2.9^{+0.2}_{-0.2}$&$0.31^{+0.06}_{-0.06}$\\
NGC 5927&$4.2^{+0.0}_{-0.0}$&$5.5^{+0.1}_{-0.1}$&$0.13^{+0.01}_{-0.01}$&NGC 6584&$2.4^{+0.2}_{-0.2}$&$24.8^{+1.5}_{-1.5}$&$0.82^{+0.0}_{-0.0}$\\
NGC 5946&$0.7^{+0.2}_{-0.1}$&$5.5^{+0.4}_{-0.3}$&$0.76^{+0.02}_{-0.03}$&Mercer 5&$2.0^{+0.1}_{-0.0}$&$5.7^{+0.6}_{-0.5}$&$0.49^{+0.02}_{-0.03}$\\
BH 176&$12.4^{+1.0}_{-1.0}$&$21.5^{+8.3}_{-5.1}$&$0.27^{+0.11}_{-0.1}$&NGC 6624&$0.2^{+0.0}_{-0.0}$&$1.3^{+0.0}_{-0.0}$&$0.69^{+0.03}_{-0.04}$\\
NGC 5986&$0.4^{+0.1}_{-0.1}$&$5.1^{+0.1}_{-0.1}$&$0.86^{+0.02}_{-0.02}$&NGC 6626&$0.3^{+0.0}_{-0.0}$&$3.0^{+0.1}_{-0.1}$&$0.8^{+0.01}_{-0.01}$\\
FSR 1716&$2.4^{+0.2}_{-0.2}$&$5.1^{+0.2}_{-0.1}$&$0.36^{+0.02}_{-0.02}$&NGC 6638&$0.1^{+0.0}_{-0.0}$&$2.3^{+0.2}_{-0.3}$&$0.93^{+0.01}_{-0.01}$\\
Pal 14&$1.3^{+1.4}_{-0.8}$&$118.6^{+2.7}_{-2.5}$&$0.98^{+0.01}_{-0.02}$&NGC 6637&$0.7^{+0.1}_{-0.1}$&$1.8^{+0.1}_{-0.0}$&$0.42^{+0.04}_{-0.04}$\\
BH 184&$1.7^{+0.1}_{-0.1}$&$4.6^{+0.1}_{-0.1}$&$0.46^{+0.01}_{-0.02}$&NGC 6642&$0.1^{+0.0}_{-0.0}$&$1.9^{+0.0}_{-0.0}$&$0.91^{+0.03}_{-0.04}$\\
NGC 6093&$1.4^{+0.1}_{-0.1}$&$4.1^{+0.1}_{-0.1}$&$0.49^{+0.02}_{-0.02}$&NGC 6652&$0.3^{+0.1}_{-0.1}$&$3.0^{+0.1}_{-0.1}$&$0.82^{+0.04}_{-0.04}$\\
NGC 6121&$0.4^{+0.0}_{-0.0}$&$6.6^{+0.0}_{-0.0}$&$0.88^{+0.01}_{-0.01}$&NGC 6656&$3.0^{+0.0}_{-0.0}$&$9.9^{+0.0}_{-0.0}$&$0.54^{+0.0}_{-0.0}$\\
NGC 6101&$10.1^{+0.1}_{-0.1}$&$45.7^{+1.6}_{-1.4}$&$0.64^{+0.01}_{-0.01}$&Pal 8&$1.3^{+0.5}_{-0.4}$&$4.1^{+0.6}_{-0.5}$&$0.53^{+0.07}_{-0.07}$\\
NGC 6144&$2.2^{+0.1}_{-0.1}$&$3.2^{+0.0}_{-0.0}$&$0.19^{+0.02}_{-0.02}$&NGC 6681&$0.4^{+0.1}_{-0.1}$&$5.2^{+0.2}_{-0.2}$&$0.85^{+0.02}_{-0.03}$\\
NGC 6139&$1.8^{+0.3}_{-0.3}$&$3.8^{+0.3}_{-0.3}$&$0.37^{+0.03}_{-0.03}$&NGC 6712&$1.2^{+0.1}_{-0.1}$&$4.7^{+0.0}_{-0.0}$&$0.6^{+0.02}_{-0.02}$\\
Terzan 3&$2.3^{+0.1}_{-0.1}$&$2.9^{+0.1}_{-0.0}$&$0.11^{+0.01}_{-0.0}$&NGC 6715&$15.1^{+0.4}_{-0.4}$&$62.6^{+4.4}_{-4.1}$&$0.61^{+0.01}_{-0.01}$\\
NGC 6171&$1.4^{+0.0}_{-0.0}$&$3.8^{+0.0}_{-0.0}$&$0.45^{+0.01}_{-0.01}$&NGC 6717&$0.9^{+0.0}_{-0.0}$&$2.4^{+0.0}_{-0.0}$&$0.45^{+0.0}_{-0.01}$\\
ESO 452-11&$0.3^{+0.0}_{-0.0}$&$2.3^{+0.1}_{-0.0}$&$0.76^{+0.02}_{-0.02}$&NGC 6723&$1.9^{+0.0}_{-0.0}$&$2.6^{+0.0}_{-0.0}$&$0.16^{+0.01}_{-0.01}$\\
NGC 6205&$2.3^{+0.0}_{-0.0}$&$8.8^{+0.0}_{-0.0}$&$0.58^{+0.0}_{-0.0}$&NGC 6749&$1.4^{+0.1}_{-0.1}$&$5.0^{+0.1}_{-0.0}$&$0.57^{+0.01}_{-0.01}$\\
NGC 6229&$2.2^{+0.3}_{-0.3}$&$30.2^{+0.4}_{-0.4}$&$0.86^{+0.02}_{-0.02}$&NGC 6752&$3.4^{+0.0}_{-0.0}$&$5.4^{+0.0}_{-0.0}$&$0.24^{+0.0}_{-0.0}$\\
NGC 6218&$2.3^{+0.0}_{-0.0}$&$4.7^{+0.0}_{-0.0}$&$0.35^{+0.0}_{-0.0}$&NGC 6760&$1.8^{+0.1}_{-0.0}$&$5.7^{+0.2}_{-0.2}$&$0.51^{+0.02}_{-0.02}$\\
FSR 1735&$1.2^{+0.1}_{-0.1}$&$4.0^{+0.3}_{-0.3}$&$0.53^{+0.03}_{-0.05}$&NGC 6779&$1.9^{+0.1}_{-0.1}$&$13.1^{+0.2}_{-0.2}$&$0.74^{+0.01}_{-0.01}$\\
NGC 6235&$3.9^{+0.4}_{-0.4}$&$9.0^{+1.6}_{-1.4}$&$0.39^{+0.03}_{-0.03}$&Terzan 7&$14.8^{+0.5}_{-0.5}$&$73.8^{+10.2}_{-8.7}$&$0.67^{+0.03}_{-0.03}$\\
NGC 6254&$1.5^{+0.0}_{-0.0}$&$4.6^{+0.0}_{-0.0}$&$0.5^{+0.0}_{-0.0}$&Pal 10&$6.4^{+1.1}_{-1.1}$&$12.8^{+7.1}_{-3.4}$&$0.33^{+0.12}_{-0.05}$\\
NGC 6256&$1.3^{+0.2}_{-0.2}$&$2.1^{+0.0}_{-0.0}$&$0.21^{+0.08}_{-0.07}$&Arp 2&$18.4^{+0.4}_{-0.4}$&$73.9^{+6.0}_{-5.1}$&$0.6^{+0.02}_{-0.02}$\\
Pal 15&$2.0^{+0.7}_{-0.7}$&$52.7^{+1.4}_{-1.2}$&$0.93^{+0.02}_{-0.02}$&NGC 6809&$1.6^{+0.0}_{-0.0}$&$5.8^{+0.0}_{-0.0}$&$0.56^{+0.0}_{-0.01}$\\
NGC 6266&$0.9^{+0.0}_{-0.0}$&$2.3^{+0.1}_{-0.1}$&$0.45^{+0.01}_{-0.01}$&Terzan 8&$18.0^{+0.4}_{-0.4}$&$92.2^{+9.8}_{-8.6}$&$0.67^{+0.02}_{-0.02}$\\
NGC 6273&$1.3^{+0.0}_{-0.0}$&$3.4^{+0.1}_{-0.1}$&$0.45^{+0.01}_{-0.02}$&Pal 11&$5.0^{+0.8}_{-0.7}$&$8.8^{+0.5}_{-0.4}$&$0.28^{+0.05}_{-0.05}$\\
NGC 6284&$1.7^{+0.1}_{-0.1}$&$6.7^{+0.4}_{-0.4}$&$0.59^{+0.0}_{-0.01}$&NGC 6838&$4.7^{+0.0}_{-0.0}$&$7.0^{+0.0}_{-0.0}$&$0.2^{+0.0}_{-0.0}$\\
NGC 6287&$1.3^{+0.2}_{-0.2}$&$4.6^{+0.2}_{-0.1}$&$0.57^{+0.06}_{-0.05}$&NGC 6864&$0.9^{+0.3}_{-0.3}$&$16.8^{+0.6}_{-0.5}$&$0.9^{+0.03}_{-0.03}$\\
NGC 6293&$0.2^{+0.1}_{-0.0}$&$2.4^{+0.5}_{-0.3}$&$0.84^{+0.02}_{-0.04}$&NGC 6934&$2.9^{+0.2}_{-0.2}$&$60.5^{+3.6}_{-3.2}$&$0.91^{+0.0}_{-0.0}$\\
NGC 6304&$1.4^{+0.1}_{-0.1}$&$2.6^{+0.1}_{-0.1}$&$0.3^{+0.0}_{-0.0}$&NGC 6981&$0.3^{+0.0}_{-0.0}$&$25.2^{+0.6}_{-0.4}$&$0.98^{+0.0}_{-0.0}$\\
NGC 6316&$1.1^{+0.3}_{-0.3}$&$3.9^{+0.5}_{-0.5}$&$0.56^{+0.06}_{-0.05}$&NGC 7006&$2.5^{+0.4}_{-0.4}$&$59.1^{+0.9}_{-0.9}$&$0.92^{+0.01}_{-0.01}$\\
NGC 6341&$2.1^{+0.0}_{-0.0}$&$10.6^{+0.1}_{-0.0}$&$0.67^{+0.0}_{-0.01}$&Laevens 3&$13.1^{+6.2}_{-4.9}$&$83.0^{+5.8}_{-3.8}$&$0.73^{+0.09}_{-0.09}$\\
NGC 6325&$1.1^{+0.1}_{-0.1}$&$1.3^{+0.2}_{-0.1}$&$0.07^{+0.1}_{-0.01}$&Segue 3&$26.8^{+1.6}_{-3.5}$&$29.6^{+6.8}_{-2.1}$&$0.08^{+0.07}_{-0.03}$\\
NGC 6333&$1.6^{+0.0}_{-0.0}$&$6.8^{+0.1}_{-0.1}$&$0.62^{+0.01}_{-0.01}$&NGC 7078&$4.1^{+0.1}_{-0.1}$&$10.7^{+0.1}_{-0.1}$&$0.45^{+0.01}_{-0.01}$\\
NGC 6342&$0.7^{+0.2}_{-0.1}$&$1.7^{+0.1}_{-0.0}$&$0.42^{+0.09}_{-0.1}$&NGC 7089&$1.3^{+0.1}_{-0.1}$&$19.4^{+0.3}_{-0.3}$&$0.87^{+0.01}_{-0.01}$\\
NGC 6356&$4.1^{+1.3}_{-1.0}$&$9.1^{+1.4}_{-1.2}$&$0.38^{+0.06}_{-0.06}$&NGC 7099&$1.5^{+0.0}_{-0.0}$&$8.8^{+0.1}_{-0.1}$&$0.71^{+0.0}_{-0.0}$\\
NGC 6355&$0.8^{+0.2}_{-0.1}$&$1.3^{+0.4}_{-0.1}$&$0.26^{+0.05}_{-0.02}$&Pal 12&$15.1^{+0.3}_{-0.3}$&$73.5^{+8.8}_{-7.4}$&$0.66^{+0.03}_{-0.03}$\\
NGC 6352&$2.9^{+0.0}_{-0.0}$&$3.9^{+0.1}_{-0.1}$&$0.15^{+0.0}_{-0.0}$&Pal 13&$7.1^{+0.4}_{-0.5}$&$69.4^{+3.2}_{-3.2}$&$0.82^{+0.0}_{-0.0}$\\
IC 1257&$0.8^{+0.4}_{-0.2}$&$20.0^{+1.2}_{-1.7}$&$0.92^{+0.02}_{-0.04}$&NGC 7492&$2.9^{+0.6}_{-0.5}$&$26.9^{+0.7}_{-0.7}$&$0.81^{+0.03}_{-0.03}$\\
Terzan 2&$0.2^{+0.0}_{-0.0}$&$0.8^{+0.3}_{-0.2}$&$0.65^{+0.12}_{-0.13}$&&&&

\end{longtable}


\bsp	
\label{lastpage}
\end{document}